\documentclass[a4paper,onecolumn,11p,accepted=2025-02-15]{quantumarticle}

\pdfoutput=1
\usepackage[utf8]{inputenc}
\usepackage[english]{babel}
\usepackage[T1]{fontenc}
\usepackage{hyperref}
\usepackage[numbers,sort&compress]{natbib}

\usepackage[all]{hypcap}   

\usepackage{amsmath,amssymb}
\usepackage{graphicx}
\usepackage{enumerate}
\usepackage{soul}
\usepackage{verbatim}
\usepackage{color}
\usepackage{placeins}  

\usepackage{braket}
\usepackage{dsfont}

\usepackage{tensor}
\usepackage{tikz}
\usepackage{tikz-network}
\usetikzlibrary{patterns,decorations.pathreplacing,calc}

\renewcommand{\mod}{\text{mod }}

\newcommand{\ue}{\text{e}}
\newcommand{\ui}{\text{i}}

\newcommand{\U}{\mathcal{U}}
\newcommand{\T}{\mathcal{T}}
\renewcommand{\S}{\mathcal{S}}

\newcommand{\Perm}{\mathds{P}}
\newcommand{\Proj}{\mathcal{P}}
\newcommand{\Hil}{\mathcal{H}}
\newcommand{\tr}{\text{tr}}

\newcommand{\Acompl}{{A^{\mathrm{c}}}}

\newcommand{\HIDDEN}[1]{}

\definecolor{myred}{RGB}{232,102,102}
\definecolor{myblue}{RGB}{187,187,255}
\definecolor{mygreen}{RGB}{34,139,34}
\definecolor{myorange}{RGB}{255,165,0}
\definecolor{myorange2}{RGB}{240,196,120}
\definecolor{OliveGreen}{RGB}{85,107,47}
\definecolor{NavyBlue}{RGB}{0,0,128}
\definecolor{mygray}{RGB}{184,186,194}
\definecolor{myrose}{RGB}{189, 120, 191}
\definecolor{myturquoise}{RGB}{213, 232, 235}

\newcommand{\gate}[2]{
    \draw[thick] (#1 -.5, #2 -0.5) -- (#1 + 0.5, #2 + 0.5);
    \draw[thick] (#1 -0.5, #2 + 0.5) -- (#1 + 0.5, #2 -0.5);
    \draw[thick, fill=myred, rounded corners=2pt] (#1 -0.25, #2 -0.25) rectangle  (#1 + .25,#2 + .25);
    \draw[thick] (#1 + 0.05,#2 + 0.15) -- (#1 + 0.15,#2 + 0.15) -- (#1 + 0.15, #2 + 0.05) ;
}

\newcommand{\gateadjoint}[2]{
  \draw[thick] (#1 -.5, #2 -0.5) -- (#1 + 0.5, #2 + 0.5);
    \draw[thick] (#1 -0.5, #2 + 0.5) -- (#1 + 0.5, #2 -0.5);
    \draw[thick, fill=myblue, rounded corners=2pt] (#1 -0.25, #2 -0.25) rectangle  (#1 + .25,#2 + .25);
    \draw[thick] (#1 + 0.05,#2 + 0.15) -- (#1 + 0.15,#2 + 0.15) -- (#1 + 0.15, #2 + 0.05) ;
}

\newcommand{\gatedisorder}[2]{
    \draw[thick] (#1 -.5, #2 -0.5) -- (#1 + 0.5, #2 + 0.5);
    \draw[thick] (#1 -0.5, #2 + 0.5) -- (#1 + 0.5, #2 -0.5);
    \draw[thick, fill=myred, rounded corners=2pt] (#1 -0.25, #2 -0.25) rectangle  (#1 + .25,#2 + .25);
    \draw[thick, fill=lightgreen_right] (#1 + 0.35,#2 + 0.35) circle (0.125cm);
    \draw[thick, fill=lightgreen_left] (#1 - 0.35,#2 + 0.35) circle (0.125cm);
    \draw[thick] (#1 + 0.05,#2 + 0.15) -- (#1 + 0.15,#2 + 0.15) -- (#1 + 0.15, #2 + 0.05) ;
    \draw[thick] (#1 + 0.3,#2 + 0.38) -- (#1 + 0.38,#2 + 0.38) -- (#1 + 0.38, #2 + 0.3) ;
    \draw[thick] (#1 - 0.3,#2 + 0.38) -- (#1 - 0.38,#2 + 0.38) -- (#1 - 0.38, #2 + 0.3) ;
}

\newcommand{\gatedisorderadjoint}[2]{
    \draw[thick] (#1 -.5, #2 -0.5) -- (#1 + 0.5, #2 + 0.5);
    \draw[thick] (#1 -0.5, #2 + 0.5) -- (#1 + 0.5, #2 -0.5);
    \draw[thick, fill=myblue, rounded corners=2pt] (#1 -0.25, #2 -0.25) rectangle  (#1 + .25,#2 + .25);
    \draw[thick, fill=magenta_right] (#1 + 0.35,#2 - 0.35) circle (0.125cm);
    \draw[thick, fill=magenta_left] (#1 - 0.35,#2 - 0.35) circle (0.125cm);
    \draw[thick] (#1 + 0.05,#2 + 0.15) -- (#1 + 0.15,#2 + 0.15) -- (#1 + 0.15, #2 + 0.05) ;
    \draw[thick] (#1 + 0.32,#2 - 0.4) -- (#1 + 0.32,#2 - 0.32) -- (#1 + 0.4, #2 - 0.32) ;
    \draw[thick] (#1 - 0.32,#2 - 0.4) -- (#1 - 0.32,#2 - 0.32) -- (#1 - 0.4, #2 - 0.32) ;
}

\newcommand{\dualgatedisorder}[2]{
    \draw[thick] (#1 -.5, #2 -0.5) -- (#1 + 0.5, #2 + 0.5);
    \draw[thick] (#1 -0.5, #2 + 0.5) -- (#1 + 0.5, #2 -0.5);
    \draw[thick, fill=myred, rounded corners=2pt] (#1 -0.25, #2 -0.25) rectangle  (#1 + .25,#2 + .25);
    \draw[thick, fill=lightgreen_right] (#1 - 0.35,#2 - 0.35) circle (0.125cm);
    \draw[thick, fill=lightgreen_left] (#1 + 0.35,#2 - 0.35) circle (0.125cm);
    \draw[thick] (#1 - 0.05,#2 + 0.15) -- (#1 - 0.15,#2 + 0.15) -- (#1 - 0.15, #2 + 0.05) ;
    \draw[thick] (#1 - 0.3,#2 - 0.38) -- (#1 - 0.38,#2 - 0.38) -- (#1 - 0.38, #2 - 0.3) ;
    \draw[thick] (#1 + 0.3,#2 - 0.38) -- (#1 + 0.38,#2 - 0.38) -- (#1 + 0.38, #2 - 0.3) ;
}

\newcommand{\dualgatedisorderadjoint}[2]{
    \draw[thick] (#1 -.5, #2 -0.5) -- (#1 + 0.5, #2 + 0.5);
    \draw[thick] (#1 -0.5, #2 + 0.5) -- (#1 + 0.5, #2 -0.5);
    \draw[thick, fill=myblue, rounded corners=2pt] (#1 -0.25, #2 -0.25) rectangle  (#1 + .25,#2 + .25);
    \draw[thick, fill=magenta_right] (#1 - 0.35,#2 + 0.35) circle (0.125cm);
    \draw[thick, fill=magenta_left] (#1 + 0.35,#2 + 0.35) circle (0.125cm);
    \draw[thick] (#1 + 0.05,#2 - 0.15) -- (#1 + 0.15,#2 - 0.15) -- (#1 + 0.15, #2 - 0.05) ;
    \draw[thick] (#1 - 0.32,#2 + 0.4) -- (#1 - 0.32,#2 + 0.32) -- (#1 - 0.4, #2 + 0.32) ;
    \draw[thick] (#1 + 0.32,#2 + 0.4) -- (#1 + 0.32,#2 + 0.32) -- (#1 + 0.4, #2 + 0.32) ;
}

\newcommand{\niceIdentity}[2]{
    \draw[thick] (#1 - 0.5,#2 - 0.5) -- (#1 - 0.3,#2 - 0.3) -- (#1 - 0.3, #2 + 0.3) -- (#1-0.5,#2 + 0.5);
    \draw[thick] (#1 + 0.5,#2 - 0.5) -- (#1 + 0.3,#2 - 0.3) -- (#1 + 0.3, #2 + 0.3) -- (#1+ 0.5,#2 + 0.5);
}

\newcommand{\disorder}[2]{
    \draw[thick] (#1,#2 - 0.3) -- (#1,#2 + 0.3);
    \draw[thick, fill=lightgreen] (#1,#2) circle (0.15cm);
    \draw[thick] (#1 - 0.08, #2 + 0.) -- (#1,#2 + 0.08) -- (#1 + 0.08, #2) ;
}

\newcommand{\disorderfolded}[2]{
    \draw[thick] (#1,#2 - 0.3) -- (#1,#2 + 0.3);
    \draw[thick, fill=myorange2] (#1,#2) circle (0.15cm);
    \draw[thick] (#1 - 0.08, #2 + 0.) -- (#1,#2 + 0.08) -- (#1 + 0.08, #2) ;
}

\newcommand{\disorderadjoint}[2]{
    \draw[thick] (#1,#2 - 0.3) -- (#1,#2 + 0.3);
    \draw[thick, fill=magenta] (#1,#2) circle (0.15cm);
    \draw[thick] (#1 - 0.08, #2 + 0.) -- (#1,#2 + 0.08) -- (#1 + 0.08, #2) ;
}

\newcommand{\foldedgate}[2]{
    \draw[thick] (#1 -.5, #2 -0.5) -- (#1 + 0.5, #2 + 0.5);
    \draw[thick] (#1 -0.5, #2 + 0.5) -- (#1 + 0.5, #2 -0.5);
    \draw[thick, fill=mygreen, rounded corners=2pt] (#1 -0.25, #2 -0.25) rectangle  (#1 + .25,#2 + .25);
    \draw[thick] (#1 + 0.05,#2 + 0.15) -- (#1 + 0.15,#2 + 0.15) -- (#1 + 0.15, #2 + 0.05) ;
}

\newcommand{\foldedgatedisorder}[2]{
    \draw[thick] (#1 -.5, #2 -0.5) -- (#1 + 0.5, #2 + 0.5);
    \draw[thick] (#1 -0.5, #2 + 0.5) -- (#1 + 0.5, #2 -0.5);
    \draw[thick, fill=mygreen, rounded corners=2pt] (#1 -0.25, #2 -0.25) rectangle  (#1 + .25,#2 + .25);
    \draw[thick, fill=orange_right] (#1 + 0.35,#2 - 0.35) circle (0.125cm);
    \draw[thick, fill=orange_left] (#1 - 0.35,#2 - 0.35) circle (0.125cm);
    \draw[thick] (#1 + 0.05,#2 + 0.15) -- (#1 + 0.15,#2 + 0.15) -- (#1 + 0.15, #2 + 0.05) ;
    \draw[thick] (#1 + 0.32,#2 - 0.4) -- (#1 + 0.32,#2 - 0.32) -- (#1 + 0.4, #2 - 0.32) ;
    \draw[thick] (#1 - 0.32,#2 - 0.4) -- (#1 - 0.32,#2 - 0.32) -- (#1 - 0.4, #2 - 0.32) ;
}

\newcommand{\foldedgatedisorderII}[2]{
    \draw[thick] (#1 -.5, #2 -0.5) -- (#1 + 0.5, #2 + 0.5);
    \draw[thick] (#1 -0.5, #2 + 0.5) -- (#1 + 0.5, #2 -0.5);
    \draw[thick, fill=mygreen, rounded corners=2pt] (#1 -0.25, #2 -0.25) rectangle  (#1 + .25,#2 + .25);
    \draw[thick, fill=orange_right] (#1 - 0.35,#2 + 0.35) circle (0.125cm);
    \draw[thick, fill=orange_left] (#1 - 0.35,#2 - 0.35) circle (0.125cm);
    \draw[thick] (#1 + 0.05,#2 + 0.15) -- (#1 + 0.15,#2 + 0.15) -- (#1 + 0.15, #2 + 0.05) ;
    \draw[thick] (#1 - 0.32,#2 + 0.4) -- (#1 - 0.32,#2 + 0.32) -- (#1 - 0.4, #2 + 0.32) ;
    \draw[thick] (#1 - 0.32,#2 - 0.4) -- (#1 - 0.32,#2 - 0.32) -- (#1 - 0.4, #2 - 0.32) ;
}

\newcommand{\dualgate}[2]{
    \draw[thick] (#1 -.5, #2 -0.5) -- (#1 + 0.5, #2 + 0.5);
    \draw[thick] (#1 -0.5, #2 + 0.5) -- (#1 + 0.5, #2 -0.5);
    \draw[thick, fill=myred, rounded corners=2pt] (#1 -0.25, #2 -0.25) rectangle  (#1 + .25,#2 + .25);
    \draw[thick] (#1 - 0.05,#2 + 0.15) -- (#1 - 0.15,#2 + 0.15) -- (#1 - 0.15, #2 + 0.05) ;
}

\newcommand{\dualgateadjoint}[2]{
  \draw[thick] (#1 -.5, #2 -0.5) -- (#1 + 0.5, #2 + 0.5);
    \draw[thick] (#1 -0.5, #2 + 0.5) -- (#1 + 0.5, #2 -0.5);
    \draw[thick, fill=myblue, rounded corners=2pt] (#1 -0.25, #2 -0.25) rectangle  (#1 + .25,#2 + .25);
    \draw[thick] (#1 + 0.05,#2 - 0.15) -- (#1 + 0.15,#2 - 0.15) -- (#1 + 0.15, #2 - 0.05) ;
}

\newcommand{\IdState}[2]{
    \draw[thick, fill=white] (#1,#2) circle (0.12cm);
}

\newcommand{\IdStateBottom}[2]{
    \draw[thick] (#1, #2) -- (#1, #2 + 0.3);
    \draw[thick, fill=white] (#1,#2) circle (0.12cm);
}

\begin{document}

\title{Eigenstate Correlations in Dual-Unitary Quantum Circuits: Partial Spectral Form Factor}

\author{Felix Fritzsch}
\affiliation{Max Planck Institute for the Physics of Complex Systems, 01187 Dresden, Germany}
\orcid{0000-0002-7659-7574}
\email{fritzsch@pks.mpg.de}
\author{Maximilian F.~I.~Kieler}
\orcid{0000-0002-1092-5539}
\affiliation{Technische Universit\"at Dresden,
    Institut f\"ur Theoretische Physik and Center for Dynamics,
    01062 Dresden, Germany}
\author{Arnd B\"acker}
\affiliation{Technische Universit\"at Dresden,
    Institut f\"ur Theoretische Physik and Center for Dynamics,
    01062 Dresden, Germany}
\orcid{0000-0002-4321-8099}
\maketitle

\begin{abstract}
While the notion of quantum chaos is tied to random matrix spectral 
correlations, also eigenstate properties in chaotic systems are often assumed 
to be described by random matrix theory. Analytic insights into eigenstate 
correlations can be obtained by the recently introduced partial spectral form 
factor. 
Here, we study the partial spectral form factor in chaotic dual-unitary quantum 
circuits in the thermodynamic limit. We compute the latter for a finite 
subsystem in a brickwork circuit coupled to an infinite complement.
For initial times, shorter than the subsystem's size, spatial locality and 
(dual) unitarity implies a constant partial spectral form factor, clearly 
deviating from the linear ramp of the random matrix prediction.
In contrast, for larger times we prove, that the partial spectral form factor follows the random matrix result up to exponentially suppressed corrections. 
We supplement our exact analytical results by semi-analytic computations 
performed in the thermodynamic limit as well as with numerics for finite-size 
systems.
\end{abstract}

\section{Introduction}
\label{sec:intro}

The notion of quantum chaos is intimately tied to random matrix theory \cite{Dys1962b,Wig1967,Meh2004,Haa2010,Sto2000}.
This connection is motivated by the quantum-chaos conjecture for single-particle
quantum systems with chaotic classical limit
\cite{CasValGua1980,Ber1981b,BohGiaSch1984}, which eventually was proven based
on semiclassical techniques
\cite{Gut1990,SieRic2001,Sie2002,MueHeuBraHaaAlt2004}.
In fact, random matrix spectral fluctuations have become the defining property of quantum chaotic systems.
Such random matrix spectral statistics have been observed in various many-body quantum systems, which lack a well--defined classical limit and which might exhibit additional structure, such as spatial locality, not present in the standard random matrix ensembles \cite{AleKafPolRig2016}.
More precisely, while long-range spectral statistics might show deviations from
random matrix theory, below a certain energy scale, the so-called many-body
Thouless energy, random matrix universality is typically restored. Indeed,
obtaining short range spectral statistics, e.g., the distribution of level
spacings or ratios thereof and comparing with random matrix theory has become
the standard test for a system to qualify as quantum chaotic.
While such tests usually have to be carried out numerically, analytical connections to random matrix spectral statistics can be made in terms of the spectral form factor (SFF).
The SFF is the Fourier transform of the two-point correlation function of the spectrum and hence probes correlations between pairs of levels on all energy scales \cite{Haa2010}.
It plays a major role
in semiclassically establishing random matrix spectral correlations in the single particle case
 \cite{SieRic2001,Sie2002,MueHeuBraHaaAlt2004}.
Using novel techniques in recently introduced exactly solvable models for spatially local, chaotic many-body systems, the SFF has been shown to follow random matrix theory also in systems which lack a classical limit \cite{KosLjuPro2018,BerKosPro2018,BerKosPro2021,BerKosPro2022,ChaDeCha2018b,ChaDeCha2018,GarCha2021,ChaDeCha2021,FriPro2023:p}.

The actual dynamics of many-body quantum systems, however, is not only determined by the spectral properties but strongly depends on the structure of eigenstates as well.
It is generally believed, that in chaotic and ergodic systems (most of) the
eigenstates are well described by the eigenstates of random matrices, i.e., by
random states.
While this is often a good first order approximation, notable differences are
known for the entanglement of eigenstates
\cite{VidHacBiaRig2018,VidRig2017,BeuBaeMoeHaq2018,HaqClaKha2022,HerBraBae2024:p}, their fractal dimensions \cite{BaeHaqKha2019,PauCarRodBuc2021}, the
statistics of eigenstate components \cite{SrdProSot2021} as well as for isolated
non-thermal, so-called scarred states
\cite{TurMicAbaSerPap2018,SerAbaPap2021}.
Such deviations from random matrix behavior might lead to non-universal dynamics, obstruct thermalization and hence ultimately lead to a violation of the infamous eigenstate thermalization hypothesis \cite{Sre1999,Deu1991,AleKafPolRig2016}.
Despite being of such fundamental importance, rigorous analytic results on  the structure of eigenstates in spatially local many-body quantum systems remain elusive, even within the above mentioned exactly solvable models.

The fundamental reason for the absence of exact results in those models, lies
in the fact, that properties of the stationary eigenstates are encoded in the
infinite time limit in finite systems, whereas exact solvability emerges in
either the thermodynamic limit or in the limit of a large number of local
degrees of freedom, i.e., in the semiclassical limit, respectively, while
keeping time fixed. Generically, the infinite time limit and the thermodynamic
or semiclassical limit do not commute. This problem might be partially overcome
by identifying observables, which encode at least some statistical properties
of eigenstates even at finite time.
A promising candidate for such an observable is the recently introduced partial
spectral form factor (PSFF)
\cite{GonSueSchCir2020,GarCha2021,JosElbVikVerGalZol2022,YosGarCha2023:p}.
In contrast to the SFF, the PSFF depends on both correlations in the spectrum and between the eigenstates, and hence provides a valuable tool for understanding eigenstate properties
as well as the average behavior of correlation functions between local observables \cite{GarCha2021, YosGarCha2023:p, HahLuiCha2024}.
More precisely, the PSFF is defined for a given subsystem of the total quantum system and depends on the correlations between the reduced density matrices of the eigenstates in that subsystem.
The PSFF is a time dependent quantity, whose short-time behavior is dictated by the  correlations between pairs of different eigenstates, while its long time dynamics encodes information on the entanglement of eigenstates.
It hence provides a possible route to study eigenstate entanglement from an
alternative, dynamical point of view.
Additional to their conceptual importance, there exists proposals for
experimentally measuring both the SFF and the PSFF
\cite{VasGraBarSieZol2020,JosElbVikVerGalZol2022} eventually leading to
experimental probes for both on quantum processors \cite{DonEtAl:p}. \\

In this work we study the PSFF in a class of one-dimensional local quantum circuits, which are characterized by a duality between space and time \cite{AkiWalGutGuh2016,BerKosPro2019b,ClaLam2021}.
These so-called dual-unitary circuits have emerged as one of the most fruitful solvable models for many-body quantum chaos.
For instance, they allow for exactly calculating dynamical correlation functions between local observables \cite{BerKosPro2019b,ClaLam2021,ClaLam2020,GutBraAkiWalGuh2020,PirBerCirPro2020,BerPir2020,KasPro2023,FriPro2022}, entanglement \cite{FriGhoPro2023,BerKosPro2019,RamCla2023:p,ZhoNah2020,RamRatCla2024:p} as well as local operator entanglement \cite{BerKosPro2020,BerKosPro2020b,ReiBer2021,FriGhoPro2023}. Additionally, exact results for monitored dynamics \cite{ClaHenVicLam2022,YaoCla2024:p}, eigenstate thermalization \cite{FriPro2021} and its absence \cite{LogDooPapGoo2024,HolMasPal2023:p}, as well as deep thermalization \cite{HoCho2022,ClaLam2022}, computational complexity \cite{SuzMitFuj2022}, and on the relation between dual-unitarity and conformal field theory \cite{Mas2023:p,CarTag2024:p} have been obtained.
Moreover, dual-unitary circuit dynamics has been probed experimentally
\cite{MiRouQuiEtAl,CheBohFraEtAl}.
Most importantly in the context of this paper, the SFF has been proven to follow the random matrix result at all time scales \cite{BerKosPro2021,BerKosPro2018}.
This has been achieved by exploiting the space-time duality of the circuit and by interpreting the SFF as the partition function of a two-dimensional statistical mechanics model, which can be evaluated by means of transfer operators in the spatial direction.

\begin{figure}[]
    \centering
    \includegraphics[width=12.13cm]{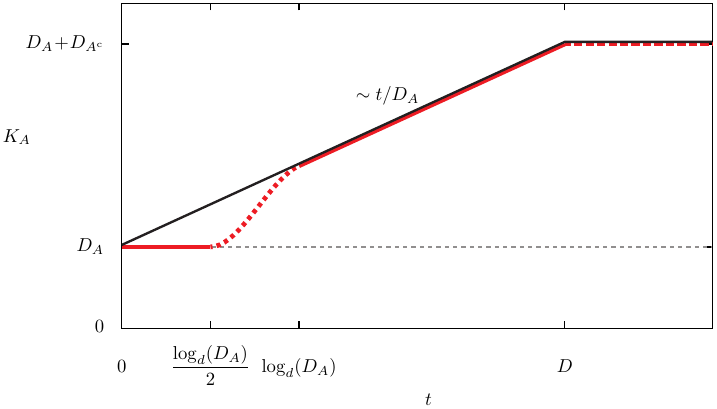}
    \caption{Schematic representation of the PSFF $K_A(t)$ for a subsystem $A$ 
of dimension $D_A$ in random matrix theory, i.e., for the CUE of dimension $D$ 
(solid black line) and our results in dual-unitary circuits (red line) with 
$d$-dimensional local Hilbert spaces. The latter include exact analytic results 
for initial times (solid red line), numerical results obtained in the 
thermodynamic limit for transient times (dotted red line), exact results for 
later times (solid red line) valid in the thermodynamic limit $D \to \infty$, as 
well as numerical data for late times $t>D$ obtained in finite systems (dashed 
red line). We indicate both the offset (thin gray dashed line) and the linear 
ramp of the PSFF. Only the leading behavior in $D_A$ and the dimension of the 
complement, $D_\Acompl$, is shown.}
    \label{fig:intro}
\end{figure}

Here we apply and extend these ideas to the computation of the PSFF in dual-unitary circuits in the thermodynamic limit.
In this limit, the PSFF for random matrices coincides with the linear ramp of the usual SFF up to a rescaling and a total constant offset, both given by the dimension of the subsystem. At times larger than the Heisenberg time, given by the inverse mean level spacing, the random matrix PSFF exhibits a plateau.
This behavior is illustrated in Fig.~\ref{fig:intro} (black line).
We demonstrate, that on short time scales, i.e., times smaller than the size of the subsystem, spatial locality implies clear deviations from the random matrix result.
This is reflected in the absence of the linear ramp and a constant PSFF, see Fig.~\ref{fig:intro} (initial solid red line).
In contrast, for sufficiently large times, we rigorously establish exponentially small bounds in time for the deviations of the PSFF in dual-unitary circuits from the random matrix result.
That is, the PSFF in dual-unitary circuits coincides with the random matrix PSFF up to exponentially small corrections (solid red line at later times).
We derive our results by constructing a tensor network representation for the PSFF resembling a two-dimensional partition function, whose boundary conditions differ between the subsystem for which the PSSF is computed and its complement.
Evaluating the latter part in the thermodynamic limit using space transfer operators, similar as for the SFF, greatly simplifies the contraction of the tensor network from being exponentially hard in time to polynomial complexity.
Due to this simplification, contracting the part of the tensor network corresponding to the subsystem can be done exactly by mapping the problem to the computation of some simple properties of just a few permutations.
While this allows for exact results, we also supplement our findings with numerical experiments, covering the regimes we cannot access analytically.
This includes both finite systems at arbitrary times, in particular for times larger than the Heisenberg time, as well as exact numerics in the thermodynamics limit on transient time scales, see Fig.~\ref{fig:intro} (dotted and dashed red line, respectively).

The remainder of this paper is organized as follows. In Sec.~\ref{sec:psff} we discuss the PSFF in more detail and highlight its relation to eigenstate properties. Subsequently, in Sec.~\ref{sec:du_circuits}, we introduce dual-unitary circuits.
Section~\ref{sec:psff_du_circuits} then contains the main part of our work in which we present the analytic computation of the PSFF.
In Sec.~\ref{sec:numerics} we supplement our analytic results with numerical experiments, before eventually concluding in Sec.~\ref{sec:outlook}.

\section{Partial Spectral Form Factor}
\label{sec:psff}

While the usual definition of quantum chaos is tied to random matrix spectral statistics, also eigenstate properties are in general believed to be well described by random matrix theory.
Eigenstate properties, however, are much harder to access by analytical methods for specific models.
A particularly simple observable is the PSFF \cite{GarCha2021,JosElbVikVerGalZol2022}, which combines correlations between eigenstates and eigenvalues.
The PSFF $K_A(t)$ is defined for a subsystem $A$ of a quantum system.
We thus consider a bipartition of the $D$-dimenisonal Hilbert space $\Hil = \Hil_A \otimes \Hil_{\Acompl}$ corresponding to the subsystem $A$ and its complement $\Acompl$, with dimension $D_A$ and $D_{\Acompl}$, respectively.
Denoting the evolution operator of the full system by $\U(t)$, we obtain the reduced evolution operator $\U_A(t)=\tr_{\Acompl}\left(\U(t)\right)$ by tracing over the degrees of freedom in the complement $\Acompl$.
Subsequently, we introduce the non-averaged PSFF as the correlator $\tr_A\left[\U_A(t)\U_A(-t)\right]$ for positive times $t>0$.
For a fixed evolution operator $\U(t)$, this quantity is heavily oscillating with time $t$ and lacks self-averaging similar to the SFF \cite{Pra1997}.
In practice such oscillations could be smoothed out by taking a moving time average of the non-averaged PSFF.
While this in principle allows to obtain the PSFF also for clean systems, for analytical computations it is advantageous to smooth out these oscillations by defining
the PSFF as the average of the above non-averaged PSFF over an ensemble of (possibly very similar/almost clean) systems.
With this, the PSFF is given as the average \cite{JosElbVikVerGalZol2022}
\begin{align}
K_A(t) = \mathds{E}\left(\tr_A\left[\U_A(t)\U_A(-t)\right]\right).
    \label{eq:psff_def2}
\end{align}
In the degenerate case when $A$ is trivial, i.e., $\Hil_A=\mathds{C}$, or equivalently, when $\Acompl$ is the whole system Eq.~\eqref{eq:psff_def2} reproduces the familiar SFF. We emphasize, that the above notation differs from Ref.~\cite{JosElbVikVerGalZol2022}, where the right hand side of Eq.~\eqref{eq:psff_def2} would have been denoted by $K_{\Acompl}(t)$. As we consider the thermodynamic limit below, by keeping $A$ fixed and letting the size of the complement $\Acompl$ tend to infinity, the notation in Eq.~\eqref{eq:psff_def2} is better suited for our purpose. \\

Given the spectral resolution $\U(t)=\sum_{n=1}^D\ue^{\ui \varphi_n t}\ket{n}\!\bra{n}$ and defining $\rho_n=\tr_{\Acompl}\left(\ket{n}\!\bra{n}\right)$ as well as $\rho_n(t)=\rho_n\ue^{\ui \varphi_n t}$ we obtain $\U_A(t) =\sum_{n=1}^D\rho_n(t)$. We hence might think of \cite{GarCha2021,HahLuiCha2024}
\begin{align}
K_A(t) = \sum_{m,n=1}^D\mathds{E}\left(\tr_A\left[\rho_n(t)\rho_m(-t)\right]\right)=\sum_{m,n=1}^D\mathds{E}\left(\tr_A\left[\rho_n\rho_m\right]\ue^{\ui (\varphi_n-\varphi_m)t}\right)
\end{align}
as the (averaged) correlator between the reduced density matrices $\rho_n(t)$ of the eigenstates $\ket{n}$. It thus encodes correlations between both eigenstates and (quasi)energies $\varphi_n$.
We emphasize, that $\rho_n(t)$ is not the time evolved $\rho_n$ as the latter is associated to an eigenstate and hence is stationary.
Assuming no degeneracies in the spectrum, one finds that the time averaged PSFF $\overline{K}_A$ yields the average purity of eigenstates (averaged over both the eigenstates and the ensemble)
as \cite{JosElbVikVerGalZol2022}.
\begin{align}
    \frac{1}{D}\overline{K}_A=\mathds{E}\left(\frac{1}{D}\sum_{n=1}^D\tr_A\left[\rho_n^2\right]\right).
    \label{eq:psff_averaged}
\end{align}
The above reasoning could be generalized by introducing moments of the partial spectral form factor as
\begin{align}
K_A^{(a,b)}(t) = \mathds{E}\left(
\left[
\tr_A\left(
\left[\U_A(t)\U_A(-t)\right]^a
\right)
\right]^b
\right)
\label{eq:psff_moments}.
 \end{align}
The moments of order $(a,1)$ encode information about higher even moments of the reduced density matrices $\rho_n^{2a}$ and consequently about the average entanglement spectrum. In contrast higher moments with $b>1$ encode information on the distribution of the moments $\rho_n^{2a}$.
In this work, however, we are exclusively concerned with the PSFF corresponding to $a=b=1$. \\

A reference point for quantum chaotic systems are the orthogonal and unitary Gaussian or circular
ensembles from random matrix theory. Here we focus on systems without anti-unitary
symmetries, e.g., systems without time-reversal invariance. Moreover, we
consider Floquet systems, describing e.g.\ periodically driven systems or
quantum circuits, and hence the relevant random matrix ensemble is the circular
unitary ensemble (CUE).
In this case the PSFF is obtained by a straightforward application of known results using the fact that the eigenvectors and eigenvalues are independent \cite{Wig1967}.
Hence the average in Eq.~\eqref{eq:psff_def2} factorizes into an average over the time-independent eigenstate correlations $\tr[\rho_n\rho_m]$ and the spectral part, which carries the whole time-dependence of the PSFF. The result
reads  \cite{JosElbVikVerGalZol2022}
\begin{align}
K_A^{\mathrm{CUE}}(t; D) = \frac{D^2}{D^2 - 1}\left(D_A - \frac{1}{D_A}\right) + \frac{D^2-D_A^{-2}}{D^2 - 1}\frac{K^{\text{CUE}}(t; D)}{D_A}
\label{eq:psff_CUE_finiteD}
\end{align}
with the CUE SFF
\begin{align}
K^{\text{CUE}}(t; D)=\min\{t, D\} = \begin{cases}
t & \quad \text{for }t < D \\
D & \quad \text{for }t \geq D
\end{cases}.
\label{eq:sff_CUE}
\end{align}
In the (thermodynamic) limit of a large system with the subsystem $A$ and hence $D_A$ kept fix, i.e., for both $D \to \infty $ and $D_{\Acompl}\to \infty$ with fixed $D/D_\Acompl=D_A$, the above becomes
\begin{align}
K_A^{\mathrm{CUE}}(t) = \lim_{D \to \infty}K_A^{\mathrm{CUE}}(t; D) = D_A + \frac{t-1}{D_A}.
\label{eq:psff_CUE}
\end{align}
As physical systems are expected to follow random matrix theory only for large systems,
we are mostly interested in the PSFF in the thermodynamic limit.

Similar as for the SFF the linear ramp $\propto t$ is a signature for chaotic systems, which is not present in integrable models.
In particular, for simple random matrix models of non-chaotic systems, characterized by a Poissonian spectrum with independent quasi-energies and hence constant SFF $K^{\mathrm{Poisson}}(t; D)=D$, we find
\begin{align}
K_A^{\mathrm{Poisson}}(t; D) = \mathrm{const.}
\end{align}
with the constant depending on the details of the statistical model used for the eigenstates.
We provide two examples in App.~\ref{app:psff_integrable}.
Hence the linear ramp of the PSFF can be seen as test for quantum chaos, similar to the SFF.

\section{Dual-Unitary Circuits}
\label{sec:du_circuits}

In this section we introduce dual-unitary circuits as prototypical class of
solvable models for many-body quantum chaos, describe the circuit design, and
specify the local gates. We moreover provide a brief review of dual unitarity
as the decisive property enabling an exact solution for the PSFF later on.

\subsection{Dual-Unitary Circuits}

We consider Floquet systems evolving in discrete time, whose evolution operators $\U$ over one time step are obtained from
one-dimensional local quantum circuits, i.e. chains of qudits of even length $2L$ subject to periodic boundary conditions.
We adopt the setup of Ref.~\cite{BerKosPro2021} but refrain from the full generality used there to simplify some of the notation.\\

The local Hilbert space  $\Hil_{\text{loc}}=\mathds{C}^d$ of each qudit is of dimension $d$, while the total Hilbert space of the chain is $\Hil=\Hil_{\text{loc}}^{\otimes 2L}=\mathds{C}^{D}$ with dimension $D=d^{2L}$.
We study two-layer brickwork circuits $\U=\U_1\U_0$ with the layers built from two-qudit gates $U_{x,x+1} \in \mathrm{U}(d^2)$ acting on neighboring sites $x$ and $x+1$ for $x \in \{1,2,\ldots,2L\}$. Here, $\mathrm{U}(d^2)$ denotes the group of unitary matrices of dimension $d^2$.
Periodic boundary conditions are imposed by identifying $2L+1$ with $1$.
The layers of the circuit are defined as
\begin{align}
    \U_0 = \bigotimes_{x=1}^{L} U_{2x-1, 2x} \quad \text{and} \quad
    \U_1 = \Pi_{2L} \bigotimes_{x=1}^{L} U_{2x, 2x+1} \Pi_{2L}^{-1}
\end{align}
with $\Pi_{2L}$ denoting the periodic shift by one site on the lattice of length $2L$. This shift is defined in the canonical product basis $\ket{j_1j_2\cdots j_{2L}}$ with $j_k \in \{1,\ldots,d\}$ by
\begin{align}
    \Pi_{2L}\ket{j_1j_2 \cdots j_{2L-1}j_{2L}}=\ket{j_{2L}j_1\cdots j_{2L-2}j_{2L-1}}.
\end{align}
The local gates $U_{x,x+1}$ are of the form
\begin{align}
    U_{x,x+1}=\left(v_x^{(0)} \otimes v_{x+1}^{(0)}\right)U_0 \text{ for odd }x \quad \text{and} \quad
    U_{x,x+1}=\left(v_x^{(1)} \otimes v_{x+1}^{(1)}\right)U_0 \text{ for even }x
\end{align}
with $U_0 \in \mathrm{U}(d^2)$ a fixed two-qudit gate and $v_x^{(i)}$ random local unitaries from $\text{U}(d)$ realizing on-site disorder.
Here, $i=0$ and $i=1$ correspond to the first layer $\U_0$ and the second layer $\U_1$ of the circuit, respectively. \\

Digrammatically, we represent the gates and their adjoints by
\begin{align}
U_{x,x+1} = \begin{tikzpicture}[baseline=-3, scale=0.7]
\pgfmathsetseed{\number\pdfrandomseed} 
\pgfmathparse{0.5*random}
\xdefinecolor{lightgreen_left}{rgb}{\pgfmathresult,0.82,0.10}
\pgfmathparse{0.5*random}
\xdefinecolor{lightgreen_right}{rgb}{\pgfmathresult,0.70,0.15}
\gatedisorder{0}{0}
\end{tikzpicture} \quad \text{and} \quad
U_{x,x+1}^\dagger = \begin{tikzpicture}[baseline=-3, scale=0.7]
\pgfmathsetseed{\number\pdfrandomseed} 
\pgfmathparse{0.5*random}
\xdefinecolor{magenta_left}{rgb}{0.7,\pgfmathresult,0.39}
\pgfmathparse{0.5*random}
\xdefinecolor{magenta_right}{rgb}{0.72,\pgfmathresult,0.41}
\gatedisorderadjoint{0}{0}
\end{tikzpicture}
\end{align}%
with the boxes representing
\begin{align}
U_0 = \begin{tikzpicture}[baseline=-3, scale=0.7]
\gate{0}{0}
\end{tikzpicture} \quad \text{and} \quad
U_0^\dagger = \begin{tikzpicture}[baseline=-3, scale=0.7]
\gateadjoint{0}{0}
\end{tikzpicture},
\end{align}%
respectively. Each wire carries the $d$-dimensional local Hilbert space $\Hil_{\mathrm{loc}}$. The small wedge indicates the orientation of the gate.
The on-site disorder is depicted as
\begin{align}
v^{(i)}_x = \begin{tikzpicture}[baseline=-3, scale=0.7]
\pgfmathsetseed{\number\pdfrandomseed} 
\pgfmathparse{0.5*random}
\xdefinecolor{lightgreen}{rgb}{\pgfmathresult,0.82,0.10}
\disorder{0}{0}
\end{tikzpicture} \quad \text{and} \quad
\left(v^{(i)}_x\right)^\dagger = \begin{tikzpicture}[baseline=-3, scale=0.7]
\pgfmathsetseed{\number\pdfrandomseed} 
\pgfmathparse{0.5*random}
\xdefinecolor{magenta}{rgb}{0.7,\pgfmathresult,0.39}
\disorderadjoint{0}{0}
\end{tikzpicture},
\end{align}%
where different shades indicate different, i.e., independent, random gates.
Using the above notation, we represent the evolution operator
as
\begin{align}
\U = \,\,\begin{tikzpicture}[baseline=7, scale=0.7]
\foreach \x in {1,...,8}
{
\foreach \y in {0, 1}
{
\pgfmathparse{0.6*abs(sin(50*\x*\x)) + 0.15}
\xdefinecolor{lightgreen_left}{rgb}{\pgfmathresult,0.82,0.10}
\pgfmathparse{0.5*abs(sin(50*\x*\x + 60*\y)) + 0.2}
\xdefinecolor{lightgreen_right}{rgb}{\pgfmathresult,0.65,0.15}
\gatedisorder{2*\x + \y}{\y}
}
}
\draw[thick] (17.8, 1.5) -- (17.5, 1.5); 
\draw[thick] (17.5, 0.5) -- (17.8, 0.5);
\draw[thick] (17.8, 1.5 ) arc (270:90: -0.1);
\draw[thick] (17.8, 0.5 ) arc (270:90: -0.1);
\foreach \x in {1,2,3}{
\draw[thick](17.8 - 0.1*\x + 0.05, 1.3) -- (17.8 - 0.1*\x - 0., 1.3);
\draw[thick](17.8 - 0.1*\x + 0.05, 0.3) -- (17.8 - 0.1*\x - 0., 0.3);
}
\draw[thick] (1.2, 0.5) -- (1.5, 0.5);
\draw[thick] (1.2, 1.5) -- (1.4, 1.5) -- (1.5, 1.6) -- (1.5, 1.7);
\draw[thick] (1.2, 1.5 ) arc (90:270: 0.1);
\draw[thick] (1.2, 0.5 ) arc (90:270: 0.1);
\foreach \x in {1,2,3}{
\draw[thick](1.2 + 0.1*\x - 0.05, 1.3) -- (1.2 + 0.1*\x + 0., 1.3);
\draw[thick](1.2 + 0.1*\x - 0.05, 0.3) -- (1.2 + 0.1*\x + 0., 0.3);
}
\foreach \x in {2,...,16}
{
\draw[thick] (\x + 0.5, 1.5 ) -- (\x + 0.5, 1.7);
}
\foreach \x in {1,...,16}
{
\draw[thick] (\x + 0.5, -0.5 ) -- (\x + 0.5, -0.7);
}
\draw[thin, gray] (1.5, -1.) -- (16.5, -1.);
\foreach \x in {1,...,16}
    {
        \draw[thin, gray] (\x + 0.5, -1.) -- (\x + 0.5, -0.85);
    }
\foreach \i in {1,...,15}
{
    \Text[x=\i + 0.5, y=-1.3 ]{\tiny $\i$};
}
\Text[x=17., y=-1.3 ]{\tiny $16=2L$};
\Text[x=11,y=-1.7]{\tiny $x$}
\end{tikzpicture}
\end{align}%
shown here for $L=8$ with periodic boundary conditions.
For later convenience we further introduce a diagrammatic representation of the folded gates \cite{BanHasVerCir2009}
\begin{align}
W_{x,x+1} = U_{x,x+1}^{\mathrm{T}} \otimes U_{x,x+1}^\dagger = \begin{tikzpicture}[baseline=-3, scale=0.7]
\xdefinecolor{orange_left}{rgb}{0.83,0.5,0.15}
\xdefinecolor{orange_right}{rgb}{0.9,0.68,0.1}
\foldedgatedisorder{0}{0}
\end{tikzpicture}  \quad \text{and} \quad
W_{0} = U_{0}^{\mathrm{T}} \otimes U_{0}^\dagger = \begin{tikzpicture}[baseline=-3, scale=0.7]
\foldedgate{0}{0}
\end{tikzpicture} \label{eq:folded_gate_disorder}
\end{align}%
with the folded disorder
\begin{align}
w_x^{(i)} = \left(v^{(i)}_x\right)^{\mathrm{T}} \otimes \left(v^{(i)}_x\right)^\dagger= \begin{tikzpicture}[baseline=-3, scale=0.7]
\xdefinecolor{lightgreen}{rgb}{0.83,0.5,0.15}
\disorderfolded{0}{0}
\end{tikzpicture}.
\end{align}%
Here, $(\cdot)^{\mathrm{T}}$ denotes transposition.
The folded picture originates from the linear vectorization mapping
$B\left(\Hil_{\mathrm{loc}}\right) \ni a = \sum_{mn}a_{mn}\ket{m}\!\bra{n}
\mapsto \sum_{mn}a_{mn}\ket{m}\otimes \ket{n} \in \Hil_{\mathrm{loc}} \otimes
\Hil_{\mathrm{loc}}$, defined using the canonical basis $\{\ket{n}|n \in
\{1,\ldots,d\}\}$. Hence each wire in the folded picture carries the
$d^2$-dimensional Hilbert space $\Hil_{\mathrm{loc}} \otimes
\Hil_{\mathrm{loc}}$. The
vectorization mapping naturally extends to an isomorphism  $B\left(\Hil\right)
\to \left(\Hil_{\mathrm{loc}} \otimes \Hil_{\mathrm{loc}}\right)^{\otimes 2L}$.
\\

To eventually give a convenient parametrization of the disorder, let us fix a basis $\{h_j|j=1,...,d^2-1\}$ of the Lie algebra  $\mathfrak{su}(d)$. We choose the $h_j$ to be Hermitian and traceless operators acting on $\Hil$, i.e, in the fundamental representation.
We then take
\begin{align}
v_x^{(i)} = \exp{\left(\ui \sum_{j=1}^{d^2 - 1} \theta_{x,j}^{(i)} h_j\right)} \label{eq:onsite_disorder}
\end{align}
with the $\theta_{x,j}^{(i)}$ independent Gaussian random variables with mean zero and variance $\sigma$. The average over those $4(d^2 -1)L$ independent random variables will facilitate the ensemble average in Eq.~\eqref{eq:psff_def2}. We might think of taking $\sigma$ arbitrary small, or more precisely of sending it to $0$ at the end of the computation, in order to describe (almost) clean systems.

\subsection{Dual-Unitary Gates}

The interactions between neighboring qudits are induced by the two-qudit gate $U_0$ and choosing it to be dual-unitary will facilitate exact solvability of the problem.
Here we give a brief overview of the main properties of dual-unitary gates.
To provide an explicit representation of such gates, let us denote by
$\sigma_z=\mathrm{diag}\left(-(d-1)/2, -(d-3)/2,\ldots, (d-1)/2\right)$ the representation of the $z$ component of the spin in the irreducible representation of $\text{SU}(2)$ on $\Hil_{\mathrm{loc}}$. For qubits, this is just the usual Pauli matrix $\sigma_z$.
Moreover, we denote by $P$ the SWAP gate acting as $P\ket{ij}=\ket{ji}$ on pairs of qudits.
Fixing arbitrary single qudit gates $u_1,u_2,u_3,u_4 \in \text{U}(d)$ different from $\mathds{1}_d$ and non-zero $J \in(0, \pi]$ we choose dual-unitary gates of the form \cite{BerKosPro2019b,ClaLam2021}
\begin{align}
    U_0 = \left(u_1 \otimes u_2\right)P\exp{(\ui J \sigma_z \otimes \sigma_z)} \left(u_3 \otimes u_4\right).
    \label{eq:du_gate}
\end{align}
For generic choices of the $u_i$ the local gates will fulfill $U_0^{\mathrm{T}}\neq U_0$ and the full circuit will not be time reversal invariant. We will consider only this case, for which the statistical properties of the corresponding quantum circuit $\U$ are expected to be described by the CUE.
 For qubits, $d=2$, it has been shown, that all dual-unitary gates are of the
form in Eq.~\eqref{eq:du_gate} \cite{BerKosPro2019b}. For larger local Hilbert
space dimension, $d>2$, however, there exist other constructions of dual-unitary
gates
\cite{ClaLam2021,ClaLamVic2023:p,RatAraLak2020,AraRatLak2021,Pro2021,
JonKheIpp2021,GomPoz2022,BorPoz2022}, which in general lead to dual-unitary
gates not captured by the parametrization~\eqref{eq:du_gate}.

With choosing $U_{0}$ dual-unitary as above, all the $U_{x,x+1}$ enjoy the property of dual unitarity as well. This means that they are unitary in both time and space direction.
Formally, let us define the dual gate $\tilde{U}_0$ by its matrix elements in the canonical product basis via
\begin{align}
    \bra{i j}\tilde{U}_0\ket{kl} = \bra{ki}U_0\ket{lj}.
\end{align}
A unitary gate for which also the dual gate is unitary is called dual-unitary.
Gates of the form~\eqref{eq:du_gate} do indeed fulfill this property \cite{BerKosPro2019b,ClaLam2021}.
Unitarity, $U_0U_0^\dagger = U_0^\dagger U_0 = \mathds{1}_{d^2}$, and, respectively, dual unitarity, $\tilde U_0\tilde U_0^\dagger = \tilde U_0^\dagger \tilde U_0 = \mathds{1}_{d^2}$, are diagrammatically expressed as
\begin{align}
\begin{tikzpicture}[baseline=12, scale=0.7]
\gate{0}{1.5}
\draw[thick] (-0.5, 0.5) -- (-0.5, 1.);
\draw[thick] (0.5, 0.5) -- (0.5, 1.);
\gateadjoint{0}{0}
\end{tikzpicture}
\, = \,
\begin{tikzpicture}[baseline=12, scale=0.7]
\draw[thick] (-0.5, 2.) -- (-0.3, 1.8) -- (-0.3, -0.3) -- (-0.5, -0.5);
\draw[thick] (0.5, 2.) -- (0.3, 1.8) -- (0.3, -0.3) -- (0.5, -0.5);
\end{tikzpicture} \, = \,
\begin{tikzpicture}[baseline=12, scale=0.7]
\gateadjoint{0}{1.5}
\draw[thick] (-0.5, 0.5) -- (-0.5, 1.);
\draw[thick] (0.5, 0.5) -- (0.5, 1.);
\gate{0}{0}
\end{tikzpicture} \quad  \text{and} \quad 
\begin{tikzpicture}[baseline=12, scale=0.7]
\dualgate{0}{1.5}
\draw[thick] (-0.5, 0.5) -- (-0.5, 1.);
\draw[thick] (0.5, 0.5) -- (0.5, 1.);
\dualgateadjoint{0}{0}
\end{tikzpicture}
\, = \,
\begin{tikzpicture}[baseline=12, scale=0.7]
\draw[thick] (-0.5, 2.) -- (-0.3, 1.8) -- (-0.3, -0.3) -- (-0.5, -0.5);
\draw[thick] (0.5, 2.) -- (0.3, 1.8) -- (0.3, -0.3) -- (0.5, -0.5);
\end{tikzpicture} \, = \,
\begin{tikzpicture}[baseline=12, scale=0.7]
\dualgateadjoint{0}{1.5}
\draw[thick] (-0.5, 0.5) -- (-0.5, 1.);
\draw[thick] (0.5, 0.5) -- (0.5, 1.);
\dualgate{0}{0}
\end{tikzpicture}\, .
\label{eq:dual_unitarity}
\end{align}%
In the folded picture this translates to unitality and dual unitality expressed as
\begin{align}
\begin{tikzpicture}[baseline=-3, scale=0.7]
\foldedgate{0}{0}
\IdState{-0.5}{-0.5}
\IdState{0.5}{-0.5}
\end{tikzpicture}\,=\,
\begin{tikzpicture}[baseline=-3, scale=0.7]
\draw[thick] (-0.25, -0.3)--(-0.25, 0.4);
\draw[thick] (+0.25, -0.3)--(+0.25, 0.4);
\IdState{-0.25}{-0.3}
\IdState{0.25}{-0.3}
\end{tikzpicture}\, , \quad
\begin{tikzpicture}[baseline=-1, scale=0.7]
\foldedgate{0}{0}
\IdState{-0.5}{0.5}
\IdState{0.5}{0.5}
\end{tikzpicture}\,=\,
\begin{tikzpicture}[baseline=-4, scale=0.7]
\draw[thick] (-0.25, +0.3)--(-0.25, -0.4);
\draw[thick] (+0.25, +0.3)--(+0.25, -0.4);
\IdState{-0.25}{0.3}
\IdState{0.25}{0.3}
\end{tikzpicture}\, , \quad
\begin{tikzpicture}[baseline=-2, scale=0.7]
\foldedgate{0}{0}
\IdState{-0.5}{0.5}
\IdState{-0.5}{-0.5}
\end{tikzpicture}\,=\,
\begin{tikzpicture}[baseline=-3, scale=0.7]
\draw[thick] (-0.3, +0.25)--(0.4, 0.25);
\draw[thick] (-0.3, -0.25)--(0.4, -0.25);
\IdState{-0.3}{0.25}
\IdState{-0.3}{-0.25}
\end{tikzpicture}\quad \text{and} \quad
\begin{tikzpicture}[baseline=-2, scale=0.7]
\foldedgate{0}{0}
\IdState{0.5}{0.5}
\IdState{0.5}{-0.5}
\end{tikzpicture}\,=\,
\begin{tikzpicture}[baseline=-3, scale=0.7]
\draw[thick] (0.3, +0.25)--(-0.4, 0.25);
\draw[thick] (0.3, -0.25)--(-0.4, -0.25);
\IdState{0.3}{0.25}
\IdState{0.3}{-0.25}
\end{tikzpicture}\, .
\label{eq:dual_unitality}
\end{align}%
Here, we introduced the notation for the vectorized and normalized identity as
\begin{align}
\begin{tikzpicture}[baseline=-1, scale=0.7]
\draw[thick] (0., 0.) -- (0., 0.5);
\IdStateBottom{0}{0}
\end{tikzpicture}\, = \frac{1}{\sqrt{d}}\ket{\mathds{1}_d}
 = \ket{\circ} \, .
\end{align}%
Note that the above properties remain valid under local unitary rotations and hence are also obeyed by the disordered gates $U_{x,x+1}$ instead of just $U_0$, leading to analogous diagrammatic rules.

\section{Partial Spectral Form Factor in Dual-Unitary Circuits}
\label{sec:psff_du_circuits}

We now turn to the computation of the partial spectral form factor. We first
develop a tensor network representation of the PSFF $K_A(t; L)$ for finite
system size $2L$  and, consequently, $D=d^{2L}$. Subsequently we contract this tensor network for either short
times or in the thermodynamic limit and larger times by means of suitable space
transfer operators yielding $K_A(t)=\lim_{L\to\infty}K_A(t; L)$.

\subsection{Transfer-Operator Approach}

In the following we obtain a convenient expression for the partial spectral form factor as two-dimensional partition function, which can be efficiently evaluated in terms of space transfer operators.
To this end, we fix the subsystem $A=\{1, \ldots, 2l-1, 2l\}$ to consist of the first $2l$ lattice sites.
The Hilbert space dimension of the subsystem is consequently $D_A = d^{2l}$.
By translational invariance under two-site shifts any connected subsystem with odd leftmost boundary gives the same result. Odd subsystem sizes could be treated in a similar fashion.
The tensor-network representation as well as the definition of the transfer operators is best derived in a diagrammatic fashion.

For a fixed realization of the circuit a diagrammatic representation of $\tr_A(\U_A(t)\U_A(-t))$, shown here for $t=3$ and $2L=16$, reads
\begin{align}
\tr_A\left(\U_A(t)\U_A(-t)\right) = \hspace{10cm} \nonumber \\
\begin{tikzpicture}[baseline=0, scale=0.7]
\draw[thin, color=mygray, fill=myturquoise, rounded corners=2pt]  (2.6, -7.8) rectangle  (8.6, 6.8);
\foreach \x in {0,...,7}
{
    \pgfmathparse{0.6*abs(sin(50*\x*\x)) + 0.15}
    \xdefinecolor{lightgreen_left}{rgb}{\pgfmathresult,0.82,0.10}
    \pgfmathparse{0.5*abs(sin(50*\x*\x)) + 0.2}
    \xdefinecolor{lightgreen_right}{rgb}{\pgfmathresult,0.65,0.15}
    \foreach \y in {0,...,2}
    {
        \gatedisorder{2*\x}{2*\y + 0.5}
    }
    \pgfmathparse{0.4*abs(sin(50*\x*\x)) + 0.1}
    \xdefinecolor{magenta_left}{rgb}{0.7,\pgfmathresult,0.39}
    \pgfmathparse{0.5*abs(sin(50*\x*\x)) + 0.2}
    \xdefinecolor{magenta_right}{rgb}{0.82,\pgfmathresult,0.41}
    \foreach \y in {0,...,2}
    {
        \gatedisorderadjoint{2*\x}{-2*\y - 1.5}
    }
}
\foreach \x in {0,...,7}
{
    \pgfmathparse{0.4*abs(sin(50*\x*\x + 30)) + 0.1}
    \xdefinecolor{lightgreen_left}{rgb}{\pgfmathresult,0.82,0.10}
    \pgfmathparse{0.5*abs(sin(50*\x*\x + 60)) + 0.2}
    \xdefinecolor{lightgreen_right}{rgb}{\pgfmathresult,0.65,0.15}
    \foreach \y in {0,...,2}
    {
        \gatedisorder{2*\x + 1}{2*\y + 1.5}
    }
    \pgfmathparse{0.4*abs(sin(50*\x*\x + 30)) + 0.1}
    \xdefinecolor{magenta_left}{rgb}{0.7,\pgfmathresult,0.39}
    \pgfmathparse{0.5*abs(sin(50*\x*\x + 60)) + 0.2}
    \xdefinecolor{magenta_right}{rgb}{0.72,\pgfmathresult,0.41}
    \foreach \y in {0,...,2}
    {
        \gatedisorderadjoint{2*\x +1}{-2*\y - 2.5}
    }
}
\foreach \y in {1,...,6}{
    \draw[thick] (15.8, \y) -- (15.5, \y); 
    \draw[thick] (15.8, \y ) arc (270:90: -0.1);
    \foreach \x in {1,2,3}{
        \draw[thick](15.8 - 0.1*\x + 0.05, \y - 0.2) -- (15.8 - 0.1*\x - 0., \y - 0.2);
    }
}
\foreach \y in {-7,...,-2}{
    \draw[thick] (15.8, \y) -- (15.5, \y); 
    \draw[thick] (15.8, \y ) arc (90:270: -0.1);
    \foreach \x in {1,2,3}{
        \draw[thick](15.8 - 0.1*\x + 0.05, \y + 0.2) -- (15.8 - 0.1*\x + 0., \y + 0.2);
    }
}
\foreach \y in {1,...,5}{
    \draw[thick] (-0.8, \y) -- (-0.5, \y); 
    \draw[thick] (-0.8, \y ) arc (90:270: 0.1);
    \foreach \x in {1,2,3}{
        \draw[thick](-0.8 + 0.1*\x - 0.05, \y - 0.2) -- (-0.8 + 0.1*\x + 0., \y - 0.2);
    }
}
\draw[thick] (-0.8, 6) -- (-0.6, 6) -- (-0.5,6.1) -- (-0.5,6.3) ; 
\draw[thick] (-0.8, 6 ) arc (90:270: 0.1);
\foreach \x in {1,2,3}{
    \draw[thick](-0.8 + 0.1*\x - 0.05, 6 - 0.2) -- (-0.8 + 0.1*\x + 0., 6 - 0.2);
}
\foreach \y in {-6,...,-2}{
    \draw[thick] (-0.8, \y) -- (-0.5, \y); 
    \draw[thick] (-0.8, \y ) arc (270:90: 0.1);
    \foreach \x in {1,2,3}{
        \draw[thick](-0.8 + 0.1*\x - 0.05, \y + 0.2) -- (-0.8 + 0.1*\x + 0., \y + 0.2);
    }
}
\draw[thick] (-0.8, -7) -- (-0.6, -7) -- (-0.5, -7.1) -- (-0.5, -7.3); 
\draw[thick] (-0.8, -7 ) arc (270:90: 0.1);
\foreach \x in {1,2,3}{
    \draw[thick](-0.8 + 0.1*\x - 0.05, -7 + 0.2) -- (-0.8 + 0.1*\x + 0., -7 + 0.2);
}
\foreach \x in {1,...,15}{
    \draw[thick] (\x - 0.5, 6.) -- (\x - 0.5, 6.3);
    \draw[thick] (\x - 0.5, -7.) -- (\x - 0.5, -7.3);
}
\foreach \x in {0,...,15}{
    \draw[thick] (\x-0.3, 6.3 ) arc (0:180: 0.1);
    \draw[thick] (\x - 0.5, -7.3 ) arc (0:180: -0.1);
    \foreach \y in {1,2,3}{
        \draw[thick](\x - 0.3, 6.35 - 0.1*\y) -- (\x - 0.3,  6.3 - 0.1*\y);
        \draw[thick](\x - 0.3, -7.35 + 0.1*\y) -- (\x - 0.3,  -7.3 + 0.1*\y);
    }
}
\foreach \x in {0,...,15}{
    \draw[thick] (\x-0.3, 6.3 ) arc (0:180: 0.1);
    \draw[thick] (\x - 0.5, -7.3 ) arc (0:180: -0.1);
    \foreach \y in {1,2,3}{
        \draw[thick](\x - 0.3, 6.35 - 0.1*\y) -- (\x - 0.3,  6.3 - 0.1*\y);
        \draw[thick](\x - 0.3, -7.35 + 0.1*\y) -- (\x - 0.3,  -7.3 + 0.1*\y);
    }
}
\foreach \x in {4,...,9}{
    \draw[thick] (\x-0.5, -1.) -- (\x-0.5, 0.); 
}
\foreach \x in {0,...,3,10,11,...,15}{
    \draw[thick] (\x-0.5, -1.) -- (\x-0.5, -0.7); 
    \draw[thick] (\x - 0.5, 0.) -- (\x-0.5, -0.3); 
    \draw[thick] (\x-0.3, -0.7 ) arc (0:180: 0.1);
    \draw[thick] (\x - 0.5, -0.3 ) arc (0:180: -0.1);
    \foreach \y in {1,2,3}{
        \draw[thick](\x - 0.3, -0.7- 0.1*\y) -- (\x - 0.3,  -0.65 - 0.1*\y);
        \draw[thick](\x - 0.3, -0.3 + 0.1*\y) -- (\x - 0.3,  -0.35 + 0.1*\y);
    }
}
\foreach \x in {-1.4}{
    \foreach \y in {-8.}{
        \draw[thick, Stealth - Stealth](\x, \y + 1.8) -- (\x, \y) -- (\x + 1.8, \y);
        \Text[x=\x + 1.95,y=\y -.25]{$x$}
        \Text[x=\x -0.32,y=\y + 1.8]{$t$}
}}
\end{tikzpicture}
\end{align}%
with the shaded region indicating the subsystem $A$ of length $2l=6$, drawn here in the bulk of the network for the sake of a better illustration. We also indicate the spatial and the time direction.
The folded picture provides a more compact representation and is obtained by folding the upper, forward time sheet of the network down on top of the lower, backward time sheet.
This results in
\begin{align}
\tr_A\left(\U_A(t)\U_A(-t)\right) = \hspace{10cm} \nonumber \\
D_A \, \, \,
\begin{tikzpicture}[baseline=58, scale=0.7]
\draw[thin, color=mygray, fill=myturquoise, rounded corners=2pt]  (2.6, -0.8) rectangle  (8.6, 6.8);
first layer
\foreach \x in {0,...,7}
{
    \pgfmathparse{0.4*abs(sin(50*\x*\x)) + 0.6}
    \xdefinecolor{orange_left}{rgb}{\pgfmathresult,0.6,0.15}
    \pgfmathparse{0.4*abs(cos(40*\x*\x*\x)) + 0.6}
    \xdefinecolor{orange_right}{rgb}{\pgfmathresult,0.5,0.1}
    \foreach \y in {0,...,2}
    {
        \foldedgatedisorder{2*\x + 1}{2*\y + 0.5}
    }
}
\foreach \x in {0,...,7}
{
    \pgfmathparse{0.5*abs(sin(50*\x*\x)) + 0.6}
    \xdefinecolor{orange_left}{rgb}{\pgfmathresult,0.4,0.2}
    \pgfmathparse{0.5*abs(cos(40*\x*\x*\x)) + 0.5}
    \xdefinecolor{orange_right}{rgb}{\pgfmathresult,0.55,0.15}
    \foreach \y in {0,...,2}
    {
        \foldedgatedisorder{2*\x}{2*\y + 1.5}
    }
}
\foreach \y in {0,...,5}{
    \draw[thick] (15.8, \y) -- (15.5, \y); 
    \draw[thick] (15.8, \y + 0.2 ) arc (270:90: -0.1);
    \foreach \x in {1,2,3}{
        \draw[thick](15.8 - 0.1*\x + 0.05, \y + 0.2) -- (15.8 - 0.1*\x - 0., \y + 0.2);
    }
}
\foreach \y in {1,...,5}{
    \draw[thick] (-0.8, \y) -- (-0.5, \y); 
    \draw[thick] (-0.8, \y + 0.2 ) arc (90:270: 0.1);
    \foreach \x in {1,2,3}{
        \draw[thick](-0.8 + 0.1*\x - 0.05, \y + 0.2) -- (-0.8 + 0.1*\x + 0., \y + 0.2);
    }
}
\draw[thick] (-0.8, 0) -- (-0.6, 0) -- (-0.5,-0.1) -- (-0.5,-0.3) ; 
\draw[thick] (-0.8, 0.2 ) arc (90:270: 0.1);
\foreach \x in {1,2,3}{
    \draw[thick](-0.8 + 0.1*\x - 0.05, 0.2) -- (-0.8 + 0.1*\x + 0., 0.2);
}
\foreach \x in {0,...,15}{
    \draw[thick] (\x - 0.5, 6.) -- (\x - 0.5, 6.3);
}
\foreach \x in {1,...,15}{
    \draw[thick] (\x - 0.5, 0.) -- (\x - 0.5, -0.3);
}
\foreach \x in {0,...,3,10,11,...,15}{
    \draw[thick] (\x-0.3, 6.3 ) arc (0:180: 0.1);
    \draw[thick] (\x - 0.5, -0.3 ) arc (0:180: -0.1);
    \foreach \y in {1,2,3}{
        \draw[thick](\x - 0.3, 6.35 - 0.1*\y) -- (\x - 0.3,  6.3 - 0.1*\y);
        \draw[thick](\x - 0.3, -0.35 + 0.1*\y) -- (\x - 0.3,  -0.3 + 0.1*\y);
    }
}
\foreach \x in {4,...,9}{
    \IdState{\x - 0.5}{6.3}
    \IdState{\x - 0.5}{-0.3}
}
\end{tikzpicture}  \, .
\label{eq:psff_folded}
\end{align}%
The factor $D_A=d^{2l}$ is a consequence of the normalization of the boundary conditions, i.e., the vectorized identity in subsystem $A$.
The latter, including the disorder average can be evaluated by switching from
evolution in time direction to evolution in spatial direction; a ubiquitous
technique in (dual-)unitary circuits \cite{LerSonAba2021,BerKloAlbLagCal2022,BerKloLu2022,BerCalColKloRyl2023,SonLerAba2021,IppKhe2021,FolZhoBer2023,KosProBer2021}.

To contract the tensor network~\eqref{eq:psff_folded} we introduce the space transfer operators $\T_t$ and $S_t$ acting on the Hilbert space $\Hil_{2t}\otimes \Hil_{2t}$.
Here $\Hil_{2t} = \Hil_{\text{loc}}^{\otimes 2t}$ is associated to the time lattice $L_t = \{1, 2, \ldots, 2t\}$ of length $2t$. They might equivalently be viewed as quantum channels
acting on $B\left(\Hil_{2t}\right)$ \cite{NieChu2010}. To be more precise, $\T_t$ is proportional to a unital quantum channel by a factor $d^2$.
We use the same notation for both transfer operators and quantum channels.
The diagrammatic representation of the transfer operators from the columns of the tensor network representation~\eqref{eq:psff_folded} reads
\begin{align}
\S_t = \mathds{E}\left( \, \, 
\begin{tikzpicture}[baseline=58, scale=0.7]
\foreach \x in {0}
{
    \xdefinecolor{orange_left}{rgb}{0.9,0.4,0.15}
    \xdefinecolor{orange_right}{rgb}{0.8,0.55,0.1}
    \foreach \y in {0,...,2}
    {
        \foldedgatedisorderII{2*\x}{2*\y + 0.5}
    }
}
\foreach \x in {0}
{
    \xdefinecolor{orange_left}{rgb}{0.95,0.5,0.15}
    \xdefinecolor{orange_right}{rgb}{0.85,0.35,0.2}
    \foreach \y in {0,...,2}
    {
        \foldedgatedisorderII{2*\x + 1}{2*\y + 1.5}
    }
}
\foreach \x in {1,2}{
    \draw[thick] (\x - 0.5, 6.) -- (\x - 0.5, 6.3);
}
\draw[thick] (1 - 0.5, 0.) -- (1 - 0.5, -0.3);
\foreach \x in {1,2}{
    \draw[thick] (\x-0.5, 6.3 ) arc (0:180: 0.1);
    \foreach \y in {1,2,3}{
        \draw[thick](\x - 0.7, 6.35 - 0.1*\y) -- (\x - 0.7,  6.3 - 0.1*\y);
    }
}
\draw[thick] (1 - 0.7, -0.3 ) arc (0:180: -0.1);
\foreach \y in {1,2,3}{
    \draw[thick](1 - 0.7, -0.35 + 0.1*\y) -- (1 - 0.7,  -0.3 + 0.1*\y);
}
\draw[thick] (2 - 0.7, -0.3 ) arc (0:180: -0.1);
\foreach \y in {1,2,3}{
    \draw[thick](2 - 0.7, -0.35 + 0.1*\y) -- (2 - 0.7,  -0.3 + 0.1*\y);
}
\draw[thick] (1.5, -0.3) -- (1.5, -0.1) -- (1.6, 0.) -- (1.8, 0.);
\foreach \y in {0,...,5}{
    \draw[thick] (-0.7, \y) -- (- 0.5, \y);
}
\foreach \y in {1,...,5}{
    \draw[thick] (1.5, \y) -- (1.8, \y);
}
\end{tikzpicture}  \right)
\qquad \text{and} \qquad
\T_t = d^2 \, \,  
\begin{tikzpicture}[baseline=58, scale=0.7]
\foreach \x in {0}
{
    \foreach \y in {0,...,2}
    {
        \foldedgate{2*\x}{2*\y + 0.5}
    }
}
\foreach \x in {0}
{
    \foreach \y in {0,...,2}
    {
        \foldedgate{2*\x + 1}{2*\y + 1.5}
    }
}
\draw[thick] (1 - 0.5, 0.) -- (1 - 0.5, -0.3);
\draw[thick] (1.5, -0.3) -- (1.5, -0.1) -- (1.6, 0.) -- (1.8, 0.);
\foreach \y in {0,...,5}{
    \draw[thick] (-0.7, \y) -- (- 0.5, \y);
}
\foreach \y in {1,...,5}{
    \draw[thick] (1.5, \y) -- (1.8, \y);
}
\foreach \x in {1,2}{
    \draw[thick] (\x - 0.5, 6.) -- (\x - 0.5, 6.3);
    \IdState{\x-0.5}{6.3}
    \IdState{\x-0.5}{-0.3}
}
\end{tikzpicture} \, \,
= d^2 \, \,  
\begin{tikzpicture}[baseline=58, scale=0.7]
\foreach \x in {0}
{
    \foreach \y in {0,...,2}
    {
        \foldedgate{2*\x}{2*\y + 0.5}
    }
}
\foreach \x in {0}
{
    \foreach \y in {0,...,1}
    {
        \foldedgate{2*\x + 1}{2*\y + 1.5}
    }
}
\draw[thick] (1 - 0.5, 0.) -- (1 - 0.5, -0.3);
\draw[thick] (0.5, 5.) -- (0.5, 5.3);
\draw[thick] (1.5, 5.3) -- (1.5, 5.1) -- (1.6, 5.) -- (1.8, 5.);
\draw[thick] (1.5, -0.3) -- (1.5, -0.1) -- (1.6, 0.) -- (1.8, 0.);
\foreach \y in {0,...,5}{
    \draw[thick] (-0.7, \y) -- (- 0.5, \y);
}
\foreach \y in {1,...,3}{
    \draw[thick] (1.5, \y) -- (1.8, \y);
}
\foreach \x in {1,2}{
    \IdState{\x-0.5}{5.3}
    \IdState{\x-0.5}{-0.3}
}
\end{tikzpicture} \, ,
\label{eq:transfer_op_def}
\end{align}%
where the unitality, Eq.~\eqref{eq:dual_unitality}, is used to simplify the expression for $\T_t$.

Formally, we construct the transfer operators from the space-time duals of the folded gates $\tilde{W}_0 = \tilde{U}_0^{\mathrm{T}} \otimes \tilde{U}_0^\dagger$ and its disordered counterpart
$\tilde{W}^\prime_{x} = \tilde{W}_0\left(w_x^{(0)}\otimes w_x^{(1)}\right)$, respectively.
Note, that for the latter we have rearranged the onsite disorder $w_x^{(i)}$ compared to Eq.~\eqref{eq:folded_gate_disorder}.
With a slight abuse of notation, we denote by $\Pi_{2t}$ the $2t$-periodic shift by one site on the time lattice $L_t = \{1,2,\ldots, 2t\}$ with local Hilbert spaces $\Hil_{\mathrm{loc}}\otimes \Hil_{\mathrm{loc}}$.
In the transfer operator $\S_t$, describing the complement $\Acompl$, we explicitly include the disorder average as
\begin{align}
\S_t = \Pi_{2t}\mathds{E}\left( \left[\tilde{W}_{x+1}^\prime\right]^{\otimes t} \right)\Pi_{2t}^{-1}\mathds{E}\left( \left[\tilde{W}_{x}^\prime\right]^{\otimes t} \right).
\label{eq:transfer_operator_B_formula}
\end{align}
As the on-site disorder on different lattice sites is independent from each other the ensemble average over Eq.~\eqref{eq:psff_folded} can be performed for each pair of columns individually as is realized for $\S_t$ above. The average is taken over the $4(d^2-1)$ random variables $\theta_{x,j}^{(i)}$ entering Eq.~\eqref{eq:onsite_disorder} and is independent from $x$.
In principle, the average can be evaluated explicitly as done in Ref.~\cite{BerKosPro2021}.
Moreover, due to the disorder being identically distributed the above average is independent of $x$.
While an individual realization of the spatial transfer operator $\S_t$ is unitary due to dual unitarity, the averaged transfer operator becomes a contraction except for only a $t$-dimensional subspace corresponding to unimodular eigenvalues, which survive the averaging.
This property will become crucial when treating the thermodynamic limit in Sec.~\ref{sec:PSFF_td_limit}.
In contrast, the transfer operator for subsystem $A$ is defined in the absence of disorder and constructed only from the gates $W_0$.
It takes the form
\begin{align}
\T_t = d^2 \left(\ket{\circ}\!\bra{\circ}\otimes\left[ \tilde{W}_{0} \right]^{\otimes t-1} \otimes \ket{\circ}\!\bra{\circ}\right)\left[\tilde{W}_{0}\right]^{\otimes t}.
\end{align}
In principle we should define $\tilde{\T}_t = \mathds{E}\left(\T_t^{(x,x+1)} \right)$ instead, where $\T_t^{(x,x+1)}$ denotes the same operator as $\T_t$ but constructed from the disordered folded gates $\tilde{W}^\prime_{x}$ and $\tilde{W}^\prime_{x+1}$, respectively.
With this definition the PSFF would read
\begin{align}
K_A(t;L) = \tr\left(\tilde{\T}_t^l\S_t^{L-l}\right).
\label{eq:psff_transferops}
\end{align}
We will, however, justify below, that the on-site disorder and the corresponding average in subsystem $A$ can be safely ignored without altering the result for the cases we are interested in.
These include both initial times $t \leq l$ for arbitrary $L$, for which we obtain
\begin{align}
K_A(t;L) = \tr\left(\T_t^l\S_t^{L-l}\right)
\label{eq:psff_transferops_short_times}
\end{align}
as well as the thermodynamic limit $L \to \infty$ for which one has
\begin{align}
K_A(t) = \lim_{L \to \infty} \tr\left(\T_t^l\S_t^{L-l}\right).
\label{eq:psff_transferops_tdl}
\end{align}
For this cases, replacing the averaged $\tilde{\T}_t$ by $\T_t$ in Eqs.~\eqref{eq:psff_transferops_short_times}~and~\eqref{eq:psff_transferops_tdl} allows for a more compact and convenient notation and simplifies some of the computations without changing the result.

\subsection{Initial Times}
\label{sec:initial_times}

For times $t \leq l$ shorter than the subsystem's size the computation of the PSFF is particularly simple, as in this case $\T_t^l$ becomes a one-dimensional operator proportional to the projection onto the normalized vectorized identity $\ket{0} = d^{-t}\ket{\mathds{1}_d}^{\otimes 2t} = d^{-t} \ket{\mathds{1}_{d^{2t}}} $.
This is best seen diagrammatically via
\begin{align}
\T_t^l = D_A \, \,  
\begin{tikzpicture}[baseline=46.5, scale=0.7]
\foreach \x in {0,...,3}
{
    \foreach \y in {0,...,2}
    {
        \foldedgate{2*\x}{2*\y + 0.5}
    }
}
\foreach \x in {0,...,3}
{
    \foreach \y in {0,...,1}
    {
        \foldedgate{2*\x + 1}{2*\y + 1.5}
    }
}
\foreach \x in {1,...,7}{
\draw[thick] (\x - 0.5, 0.) -- (\x - 0.5, -0.3);
\draw[thick] (\x - 0.5, 5.) -- (\x - 0.5, 5.3);   
}
\draw[thick] (7.5, 5.3) -- (7.5, 5.1) -- (7.6, 5.) -- (7.8, 5.);
\draw[thick] (7.5, -0.3) -- (7.5, -0.1) -- (7.6, 0.) -- (7.8, 0.);
\foreach \y in {0,...,5}{
    \draw[thick] (-0.7, \y) -- (- 0.5, \y);
}
\foreach \y in {1,...,4}{
    \draw[thick] (7.5, \y) -- (7.8, \y);
}
\foreach \x in {1,...,8}{
    \IdState{\x-0.5}{5.3}
    \IdState{\x-0.5}{-0.3}
}
\end{tikzpicture}
 = D_A 
\begin{tikzpicture}[baseline=46.5, scale=0.7]
\foreach \y in {0,...,2}
{
    \foldedgate{0}{2*\y + 0.5}
}
\foreach \x in {1, 4.5}{
\foreach \y in {0,...,1}{
    \foldedgate{\x}{2*\y + 1.5}
}
}
\foreach \x in {2, 3.5}{
\foldedgate{\x}{2 + 0.5}
}
\draw[thick] (0.5, 0.) -- (0.5, -0.3);
\draw[thick] (0.5, 5.) -- (0.5, 5.3);   
\draw[thick] (5., 5.3) -- (5., 5.1) -- (5.1, 5.) -- (5.3, 5.);
\draw[thick] (5., -0.3) -- (5., -0.1) -- (5.1, 0.) -- (5.3, 0.);
\foreach \y in {0,...,5}{
    \draw[thick] (-0.7, \y) -- (- 0.5, \y);
}
\foreach \y in {1,...,4}{
    \draw[thick] (5, \y) -- (5.3, \y);
}
\foreach \x in {1,5.5}{
    \IdState{\x-0.5}{5.3}
    \IdState{\x-0.5}{-0.3}
}
\foreach \x in {2,4.5}{
    \IdState{\x-0.5}{4.}
    \IdState{\x-0.5}{1.}
}
\foreach \x in {3,3.5}{
    \IdState{\x-0.5}{3.}
    \IdState{\x-0.5}{2.}
}
\end{tikzpicture}
= D_A 
\begin{tikzpicture}[baseline=46.5, scale=0.7]
\foreach \y in {0,...,5}{
    \draw[thick] (-0.75, \y) -- (- 0.3, \y);
    \draw[thick] (0.75, \y) -- (0.3, \y);
}
\foreach \y in {0,...,5}{
    \IdState{-0.3}{\y}
    \IdState{0.3}{\y}
}
\end{tikzpicture}
\end{align}%
illustrated here for $t=3$ and $l=4$.
In the first equality we used the unitality, Eq.~\eqref{eq:dual_unitality}, of the folded gate in both forward and backward time direction. This already shows, that
$\T_t^l=D_A\ket{L}\!\bra{R}$ is one-dimensional for any unitary circuit and consequently $K_A(t; L) = D_A\bra{R}\S_t^{L-l}\ket{L}$.
For dual-unitary gates unitality in the spatial directions, 
Eq.~\eqref{eq:dual_unitality}, allows for explicitly computing 
$\ket{L}=\ket{R}=\ket{0}$.
We emphasize, that the above computation only depends on the dual unitarity of the local gates. The same result could be obtained by including the on-site disorder, with the final result not depending on the disorder realization. Consequently, the disorder average over subsystem $A$ would be trivial. This justifies our previous claim, that we can ignore the disorder in subsystem $A$ completely, without altering the result.
With a similar computation as above, one additionally finds, that $\T_t\ket{0}=d^2\ket{0}$, i.e. $\ket{0}$ is the only eigenstate for non-zero eigenvalue. In general, however, $\T_t$ is not diagonalizable, but exhibits non-trivial Jordan blocks of size at most $t$. Hence, despite having only eigenvalues $d^2$ and $0$ it is not proportional to a projection operator.

Dual unitarity of the gates in the presence of disorder further implies
\begin{align}
\S_t\ket{0} = \, \,
\begin{tikzpicture}[baseline=58, scale=0.7]
\foreach \y in {0,...,5}{
    \draw[thick] (0.75, \y) -- ( 0.3, \y);
    \IdState{0.3}{\y}
}
\end{tikzpicture} \, \,
\mathds{E}\left( \, \, 
\begin{tikzpicture}[baseline=58, scale=0.7]
\foreach \x in {0}
{
        \xdefinecolor{orange_left}{rgb}{0.9,0.4,0.15}
    \xdefinecolor{orange_right}{rgb}{0.8,0.55,0.1}<
    \foreach \y in {0,...,2}
    {
        \foldedgatedisorderII{2*\x}{2*\y + 0.5}
    }
}
\foreach \x in {0}
{
        \xdefinecolor{orange_left}{rgb}{0.95,0.5,0.15}
    \xdefinecolor{orange_right}{rgb}{0.85,0.35,0.2}
    \foreach \y in {0,...,2}
    {
        \foldedgatedisorderII{2*\x + 1}{2*\y + 1.5}
    }
}
\foreach \x in {1,2}{
    \draw[thick] (\x - 0.5, 6.) -- (\x - 0.5, 6.3);
}
\draw[thick] (1 - 0.5, 0.) -- (1 - 0.5, -0.3);
\foreach \x in {1,2}{
    \draw[thick] (\x-0.5, 6.3 ) arc (0:180: 0.1);
    \foreach \y in {1,2,3}{
        \draw[thick](\x - 0.7, 6.35 - 0.1*\y) -- (\x - 0.7,  6.3 - 0.1*\y);
    }
}
\draw[thick] (1 - 0.7, -0.3 ) arc (0:180: -0.1);
\foreach \y in {1,2,3}{
    \draw[thick](1 - 0.7, -0.35 + 0.1*\y) -- (1 - 0.7,  -0.3 + 0.1*\y);
}
\draw[thick] (2 - 0.7, -0.3 ) arc (0:180: -0.1);
\foreach \y in {1,2,3}{
    \draw[thick](2 - 0.7, -0.35 + 0.1*\y) -- (2 - 0.7,  -0.3 + 0.1*\y);
}
\draw[thick] (1.5, -0.3) -- (1.5, -0.1) -- (1.6, 0.) -- (1.8, 0.);
\foreach \y in {0,...,5}{
    \draw[thick] (-0.7, \y) -- (- 0.5, \y);
}
\foreach \y in {1,...,5}{
    \draw[thick] (1.5, \y) -- (1.8, \y);
}
\end{tikzpicture}  
\right) \, \, = \mathds{E}\left( \, \, 
\begin{tikzpicture}[baseline=58, scale=0.7]
\foreach \x in {0}
{
    \xdefinecolor{orange_left}{rgb}{0.9,0.4,0.15}
\xdefinecolor{orange_right}{rgb}{0.8,0.55,0.1}
    \foreach \y in {0,...,2}
    {
        \foldedgatedisorderII{2*\x}{2*\y + 0.5}
    }
}
\foreach \x in {0}
{
       \xdefinecolor{orange_left}{rgb}{0.95,0.5,0.15}
   \xdefinecolor{orange_right}{rgb}{0.85,0.35,0.2}
    \foreach \y in {0,...,2}
    {
        \foldedgatedisorderII{2*\x + 1}{2*\y + 1.5}
    }
}
\foreach \x in {1,2}{
    \draw[thick] (\x - 0.5, 6.) -- (\x - 0.5, 6.3);
}
\draw[thick] (1 - 0.5, 0.) -- (1 - 0.5, -0.3);
\foreach \x in {1,2}{
    \draw[thick] (\x-0.5, 6.3 ) arc (0:180: 0.1);
    \foreach \y in {1,2,3}{
        \draw[thick](\x - 0.7, 6.35 - 0.1*\y) -- (\x - 0.7,  6.3 - 0.1*\y);
    }
}
\draw[thick] (1 - 0.7, -0.3 ) arc (0:180: -0.1);
\foreach \y in {1,2,3}{
    \draw[thick](1 - 0.7, -0.35 + 0.1*\y) -- (1 - 0.7,  -0.3 + 0.1*\y);
}
\draw[thick] (2 - 0.7, -0.3 ) arc (0:180: -0.1);
\foreach \y in {1,2,3}{
    \draw[thick](2 - 0.7, -0.35 + 0.1*\y) -- (2 - 0.7,  -0.3 + 0.1*\y);
}
\draw[thick] (1.5, -0.3) -- (1.5, -0.1) -- (1.6, 0.) -- (1.8, 0.);
\foreach \y in {0,...,5}{
    \draw[thick] (-0.7, \y) -- (- 0.5, \y);
}
\foreach \y in {1,...,5}{
    \draw[thick] (1.5, \y) -- (1.8, \y);
}
\foreach \y in {0,...,5}{
    \IdState{-0.8}{\y}
}
\end{tikzpicture}  \right) \, \, = \mathds{E}\left( \, \,
\begin{tikzpicture}[baseline=46.5, scale=0.7]
\foreach \y in {0,...,5}{
    \draw[thick] (0.7, \y) -- ( 0.3, \y);
    \IdState{0.3}{\y}
}
\end{tikzpicture} \, \, \right) = \ket{0}
\end{align}%
i.e., $\ket{0}$ is also an eigenvector for $\S_t$ with eigenvalue 1. Alternatively, this can also be seen in non-vectorized form as sketched in Eq.~\eqref{eq:transfer_operator_eigenvectors} below.
Hence
\begin{align}
    K_A(t; L) = D_A \bra{0} \S_t^{L-l}\ket{0} = D_A
    \label{eq:constant_psff_initial_times}
\end{align}
is constant for initial times. This is a consequence of locality and unitarity of the evolution operator $\U$ and in contrast to the random matrix result. Note that this result is true both for finite systems and in the thermodynamic limit $L \to \infty$ at fixed subsystem size $l$ and fixed time $t \leq l$.
These notable deviation from the linear ramp of the PSFF in random matrix theory given by Eq.~\eqref{eq:psff_CUE} is our first main result.

\subsection{Partial Spectral Form Factor in the Thermodynamic Limit}
\label{sec:PSFF_td_limit}

In this section we demonstrate, that the PSFF in the termodynamic limit $L \to \infty$ at fixed subsystem size $l$ and fixed time $t > 2l$ is exponentially close to the random matrix result~\eqref{eq:psff_CUE}.
Our results are based on reducing the problem of $\S_t$ and $\T_t$ acting on a $d^{4t}$ dimensional space to a problem on a set of
$t$ permutations, thereby drastically reducing the complexity of the problem. \\

To this end we first introduce some useful notation. Let us denote the unitary representation of the permuation group $S_{2t}$ which permutes tensor factors in $\Hil_{2t}$ by $\mathds{P}$. We think of $\pi \in S_{2t} = S_{L_t}$ as a permutation of the lattice sites in $L_t = \{1, 2, \ldots, 2t\}$. The corresponding unitary operator acts on the product basis according to
\begin{align}
    \mathds{P}_{\pi}\ket{i_1i_2\cdots i_{2t}} = \ket{i_{\pi^{-1}(1)}i_{\pi^{-1}(2)}\cdots i_{\pi^{-1}(2t)}}.
\end{align}
Let us in particular denote by $\eta \in S_{2t}$ the periodic shift $\eta(i) = (i + 1)\,\mod 2t$. That is, $\mathds{P}_{\eta} = \Pi_{2t}$, which enters Eq.~\eqref{eq:transfer_operator_B_formula}.
The permutation operators have Hilbert-Schmidt norm $\|\mathds{P}_\pi \|=d^{t}$ and we  define $\ket{\pi}=d^{-t}\ket{\mathds{P}_\pi}$ as their normalized vectorization.
For a pair of permutations $\pi, \mu \in S_{2t}$ one has \cite{Col2003}
\begin{align}
    \braket{\pi | \mu} = d^{-2t} \tr\left(\mathds{P}_\pi^{\dagger}\mathds{P}_\mu\right) = d^{-2t + |\pi^{-1}\mu|},
    \label{eq:overlapp_permutations}
\end{align}
where $|\pi|$ denotes the number of disjoint cycles in the cycle decomposition of $\pi \in S_{2t}$. Note that $\braket{\pi| \mu} = \braket{\mathrm{id} | \pi^{-1}\mu}$.
We  refer to $\pi \in S_{L_t}$, the unitary operator $\mathds{P}_\pi$ as well as its normalized vectorization $\ket{\pi}$ as permutation. Moreover, we abuse this notation and terminology when referring to permutations $\pi \in S_K$ which act only on a sublattice $K \subseteq L_t$. In any case, the corresponding interpretation will be clear from the respective context. \\

We now use this notation for an efficient computation of the PSFF in the thermodynamic limit.
We start from the representation~\eqref{eq:psff_transferops_tdl} and employ, that in the limit $L \to \infty$ the operator $\S_t^{L-l}$ is determined by the leading part of the spectrum of $\S_t$ and the corresponding left and right eigenvectors.
This strategy has been employed for computing the SFF in dual unitary circuits in Ref.~\cite{BerKosPro2021}. There, a full characterization of the relevant leading part of the spectrum was proven.
In particular, it was shown that without taking the disorder average, the transfer operator would be unitary as a consequence of dual unitarity and hence all eigenvalues would have modulus one.
In contrast, upon averaging, most eigenvalues are pushed inside the unit disk and only those eigenvalues, for which the corresponding eigenvectors are invariant under the cyclic time-translation symmetry, remain unchanged.
More precisely, the results of Ref.~\cite{BerKosPro2021} imply,
that in our setting, the leading eigenvalue of $\S_t$ is 1, that its algebraic and geometric multiplicity is $t$, and that the corresponding left and right eigenvectors are given by the even shifts $\ket{r}=\ket{\eta^{2r}}$ for $r \in \{0,1,\ldots,t-1\}$.
That these are indeed eigenvectors for eigenvalue $1$ is best seen when viewing $\S_t$ as a quantum channel.
To do so, we represent $\Perm_{\eta^{2r}}$ diagrammatically, here for $t=3$ and $r=1$, as
\begin{align}
\Perm_{\eta^{2r}} = \, \,
\begin{tikzpicture}[baseline=-9, scale=0.7]
 \foreach \x in {0,...,3}{
    \draw[thick, rounded corners=2pt] (\x - 0.5, -1.) -- (\x - 0.5, -0.75)  -- (\x + 2 - 0.5, 0.25) --  (\x + 2 - 0.5, 0.5);
}
\draw[thick, rounded corners=2pt] (4 - 0.5, -1.) -- (4 - 0.5, -0.75)  -- (4 + 1.5 - 0.5, 0.)  -- (4 + 1.5 - 0.2, 0.);
\draw[thick, rounded corners=2pt] (5 - 0.5, -1.) -- (5 - 0.5, -0.75)  -- (5 + .5 - 0.5, -0.5)  -- (5 + .5 - 0.2, -0.5);
\draw[thick, rounded corners=2pt] (- 0.5, 0.5) -- (- 0.5, 0.25)  -- (-1., 0.)  -- (-1.3, 0.);
\draw[thick, rounded corners=2pt] (0.5, 0.5) -- (0.5, 0.25)  -- (-1., -0.5)  -- (-1.3, -0.5);
\draw[thick] (-1.3, -0.7) arc (270:90: 0.1);
\draw[thick] (-1.3, -0.) arc (270:90: 0.1);
\draw[thick] (5.3, -0.7) arc (90:270: -0.1);
\draw[thick] (5.3, -0.) arc (90:270: -0.1);
\foreach \x in {1,2,3}{
    \draw[thick](-1.3 + 0.1*\x - 0.05, -0.7) -- (-1.3 + 0.1*\x + 0., -0.7);
    \draw[thick](-1.3 + 0.1*\x - 0.05, 0.2) -- (-1.3 + 0.1*\x + 0., 0.2);
    \draw[thick](5.3 - 0.1*\x + 0.05, -0.7) -- (5.3 - 0.1*\x + 0., -0.7);
    \draw[thick](5.3 - 0.1*\x + 0.05, 0.2) -- (5.3 - 0.1*\x + 0., 0.2);
}
\end{tikzpicture}
\end{align}%
and compute for a fixed disorder realization (with space and time rotated by $90^\circ$)
\begin{align}
\begin{tikzpicture}[baseline=-9, scale=0.7]
\xdefinecolor{lightgreen_left}{rgb}{0.6,0.82,0.10}
\xdefinecolor{lightgreen_right}{rgb}{0.35,0.70,0.15}
\xdefinecolor{magenta_left}{rgb}{0.7,0.3,0.39}
\xdefinecolor{magenta_right}{rgb}{0.72,0.5,0.41}
\foreach \x in {0,...,2}
{
    \dualgatedisorder{2*\x}{1}
    \dualgatedisorderadjoint{2*\x}{- 1.5}
}
\xdefinecolor{lightgreen_left}{rgb}{0.25,0.82,0.10}
\xdefinecolor{lightgreen_right}{rgb}{0.5,0.70,0.15}
\xdefinecolor{magenta_left}{rgb}{0.85,0.3,0.39}
\xdefinecolor{magenta_right}{rgb}{0.72,0.1,0.41}
\foreach \x in {0,...,2}
{
    \dualgatedisorder{2*\x + 1}{+ 2.}
    \dualgatedisorderadjoint{2*\x +1}{ - 2.5}
}
\foreach \y in {1.5,2.5}{
    \draw[thick] (5.8, \y) -- (5.5, \y); 
    \draw[thick] (5.8, \y ) arc (270:90: -0.1);
    \foreach \x in {1,2,3}{
        \draw[thick](5.8 - 0.1*\x + 0.05, \y - 0.2) -- (5.8 - 0.1*\x - 0., \y - 0.2);
    }
}
\foreach \y in {-2,-3}{
    \draw[thick] (5.8, \y) -- (5.5, \y); 
    \draw[thick] (5.8, \y ) arc (90:270: -0.1);
    \foreach \x in {1,2,3}{
        \draw[thick](5.8 - 0.1*\x + 0.05, \y + 0.2) -- (5.8 - 0.1*\x + 0., \y + 0.2);
    }
}
\draw[thick] (-0.8, 1.5) -- (-0.5, 1.5); 
\draw[thick] (-0.8, 1.5 ) arc (90:270: 0.1);
\foreach \x in {1,2,3}{
    \draw[thick](-0.8 + 0.1*\x - 0.05, 1.5 - 0.2) -- (-0.8 + 0.1*\x + 0., 1.5 - 0.2);
}
\draw[thick] (-0.8, 2.5) -- (-0.6, 2.5) -- (-0.5,2.6) -- (-0.5,2.8) ; 
\draw[thick] (-0.8, 2.5 ) arc (90:270: 0.1);
\foreach \x in {1,2,3}{
    \draw[thick](-0.8 + 0.1*\x - 0.05, 2.5 - 0.2) -- (-0.8 + 0.1*\x + 0., 2.5 - 0.2);
}
\draw[thick] (-0.8, -2) -- (-0.5, -2); 
\draw[thick] (-0.8, -3) -- (-0.6, -3) -- (-0.5, -3.1) -- (-0.5, -3.3); 
\foreach \y in {-2, -3}{
    \draw[thick] (-0.8, \y ) arc (270:90: 0.1);
    \foreach \x in {1,2,3}{
        \draw[thick](-0.8 + 0.1*\x - 0.05, \y  + 0.2) -- (-0.8 + 0.1*\x + 0., \y  + 0.2);
    }
}
\foreach \x in {1,...,5}{
    \draw[thick] (\x - 0.5, 2.5) -- (\x - 0.5, 2.8);
    \draw[thick] (\x - 0.5, -3.) -- (\x - 0.5, -3.3);
}
\foreach \x in {0,...,3}{
    \draw[thick, rounded corners=2pt] (\x - 0.5, -1.) -- (\x - 0.5, -0.75)  -- (\x + 2 - 0.5, 0.25) --  (\x + 2 - 0.5, 0.5);
}
\draw[thick, rounded corners=2pt] (4 - 0.5, -1.) -- (4 - 0.5, -0.75)  -- (4 + 1.5 - 0.5, 0.)  -- (4 + 1.5 - 0.2, 0.);
\draw[thick, rounded corners=2pt] (5 - 0.5, -1.) -- (5 - 0.5, -0.75)  -- (5 + .5 - 0.5, -0.5)  -- (5 + .5 - 0.2, -0.5);
\draw[thick, rounded corners=2pt] (- 0.5, 0.5) -- (- 0.5, 0.25)  -- (-1., 0.)  -- (-1.3, 0.);
\draw[thick, rounded corners=2pt] (0.5, 0.5) -- (0.5, 0.25)  -- (-1., -0.5)  -- (-1.3, -0.5);
\draw[thick] (-1.3, -0.7) arc (270:90: 0.1);
\draw[thick] (-1.3, -0.) arc (270:90: 0.1);
\draw[thick] (5.3, -0.7) arc (90:270: -0.1);
\draw[thick] (5.3, -0.) arc (90:270: -0.1);
\foreach \x in {1,2,3}{
    \draw[thick](-1.3 + 0.1*\x - 0.05, -0.7) -- (-1.3 + 0.1*\x + 0., -0.7);
    \draw[thick](-1.3 + 0.1*\x - 0.05, 0.2) -- (-1.3 + 0.1*\x + 0., 0.2);
    \draw[thick](5.3 - 0.1*\x + 0.05, -0.7) -- (5.3 - 0.1*\x + 0., -0.7);
    \draw[thick](5.3 - 0.1*\x + 0.05, 0.2) -- (5.3 - 0.1*\x + 0., 0.2);
}
\foreach \x in {-2}{
    \foreach \y in {3.3}{
        \draw[thick, Stealth - Stealth](\x, \y - 1.8) -- (\x, \y) -- (\x + 1.8, \y);
        \Text[x=\x + 1.95,y=\y + .25]{$t$}
        \Text[x=\x -0.32,y=\y - 1.8]{$x$}
}}
\end{tikzpicture} 
= 
\begin{tikzpicture}[baseline=-9, scale=0.7]
\foreach \x in {0,...,2}
{
    \niceIdentity{2*\x}{1}
    \niceIdentity{2*\x}{- 1.5}
}
\xdefinecolor{lightgreen_left}{rgb}{0.25,0.82,0.10}
\xdefinecolor{lightgreen_right}{rgb}{0.5,0.70,0.15}
\xdefinecolor{magenta_left}{rgb}{0.85,0.3,0.39}
\xdefinecolor{magenta_right}{rgb}{0.72,0.1,0.41}
\foreach \x in {0,...,2}
{
    \dualgatedisorder{2*\x + 1}{+ 2.}
    \dualgatedisorderadjoint{2*\x +1}{ - 2.5}
}
\foreach \y in {1.5,2.5}{
    \draw[thick] (5.8, \y) -- (5.5, \y); 
    \draw[thick] (5.8, \y ) arc (270:90: -0.1);
    \foreach \x in {1,2,3}{
        \draw[thick](5.8 - 0.1*\x + 0.05, \y - 0.2) -- (5.8 - 0.1*\x - 0., \y - 0.2);
    }
}
\foreach \y in {-2,-3}{
    \draw[thick] (5.8, \y) -- (5.5, \y); 
    \draw[thick] (5.8, \y ) arc (90:270: -0.1);
    \foreach \x in {1,2,3}{
        \draw[thick](5.8 - 0.1*\x + 0.05, \y + 0.2) -- (5.8 - 0.1*\x + 0., \y + 0.2);
    }
}
\draw[thick] (-0.8, 1.5) -- (-0.5, 1.5); 
\draw[thick] (-0.8, 1.5 ) arc (90:270: 0.1);
\foreach \x in {1,2,3}{
    \draw[thick](-0.8 + 0.1*\x - 0.05, 1.5 - 0.2) -- (-0.8 + 0.1*\x + 0., 1.5 - 0.2);
}
\draw[thick] (-0.8, 2.5) -- (-0.6, 2.5) -- (-0.5,2.6) -- (-0.5,2.8) ; 
\draw[thick] (-0.8, 2.5 ) arc (90:270: 0.1);
\foreach \x in {1,2,3}{
    \draw[thick](-0.8 + 0.1*\x - 0.05, 2.5 - 0.2) -- (-0.8 + 0.1*\x + 0., 2.5 - 0.2);
}
\draw[thick] (-0.8, -2) -- (-0.5, -2); 
\draw[thick] (-0.8, -3) -- (-0.6, -3) -- (-0.5, -3.1) -- (-0.5, -3.3); 
\foreach \y in {-2, -3}{
    \draw[thick] (-0.8, \y ) arc (270:90: 0.1);
    \foreach \x in {1,2,3}{
        \draw[thick](-0.8 + 0.1*\x - 0.05, \y  + 0.2) -- (-0.8 + 0.1*\x + 0., \y  + 0.2);
    }
}
\foreach \x in {1,...,5}{
    \draw[thick] (\x - 0.5, 2.5) -- (\x - 0.5, 2.8);
    \draw[thick] (\x - 0.5, -3.) -- (\x - 0.5, -3.3);
}
\foreach \x in {0,...,3}{
    \draw[thick, rounded corners=2pt] (\x - 0.5, -1.) -- (\x - 0.5, -0.75)  -- (\x + 2 - 0.5, 0.25) --  (\x + 2 - 0.5, 0.5);
}
\draw[thick, rounded corners=2pt] (4 - 0.5, -1.) -- (4 - 0.5, -0.75)  -- (4 + 1.5 - 0.5, 0.)  -- (4 + 1.5 - 0.2, 0.);
\draw[thick, rounded corners=2pt] (5 - 0.5, -1.) -- (5 - 0.5, -0.75)  -- (5 + .5 - 0.5, -0.5)  -- (5 + .5 - 0.2, -0.5);
\draw[thick, rounded corners=2pt] (- 0.5, 0.5) -- (- 0.5, 0.25)  -- (-1., 0.)  -- (-1.3, 0.);
\draw[thick, rounded corners=2pt] (0.5, 0.5) -- (0.5, 0.25)  -- (-1., -0.5)  -- (-1.3, -0.5);
\draw[thick] (-1.3, -0.7) arc (270:90: 0.1);
\draw[thick] (-1.3, -0.) arc (270:90: 0.1);
\draw[thick] (5.3, -0.7) arc (90:270: -0.1);
\draw[thick] (5.3, -0.) arc (90:270: -0.1);
\foreach \x in {1,2,3}{
    \draw[thick](-1.3 + 0.1*\x - 0.05, -0.7) -- (-1.3 + 0.1*\x + 0., -0.7);
    \draw[thick](-1.3 + 0.1*\x - 0.05, 0.2) -- (-1.3 + 0.1*\x + 0., 0.2);
    \draw[thick](5.3 - 0.1*\x + 0.05, -0.7) -- (5.3 - 0.1*\x + 0., -0.7);
    \draw[thick](5.3 - 0.1*\x + 0.05, 0.2) -- (5.3 - 0.1*\x + 0., 0.2);
}
\end{tikzpicture} 
= \Perm_{\eta^{2r}}.   
\label{eq:transfer_operator_eigenvectors}
\end{align}%
For the first equality we use dual unitarity, Eq.~\eqref{eq:dual_unitarity}, of the disordered gates, minding that the rows of the diagram correspond to fixed spatial slices  and hence the dual gates are indeed connected to their respective adjoints. Repeating this argument for the remaining rows of nontrivial gates yields the second equality.
Taking the disorder average over Eq.~\eqref{eq:transfer_operator_eigenvectors} we obtain $\S_t\ket{r} = \mathds{E}\left(\ket{r}\right)=\ket{r}$ and hence $\ket{r}$ is a right eigenvector for eigenvalue $1$.
A similar calculation, replacing the gates with their respective adjoints, shows, that $\ket{r}$ is also left eigenvector.
Further note, that $\ket{0}=d^{-t}\ket{\mathrm{id}}$ corresponds to the vectorized identity and hence is the unique eigenvector of $\T_t$ corresponding to the non-zero eigenvalue $d^2$ as discussed in Sec.~\ref{sec:initial_times}.
The remaining claims, i.e., that there are no additional unimodular eigenvectors (or even eigenvalues with larger modulus) are much harder to obtain and we refer the reader to Ref.~\cite{BerKosPro2021} for a full proof.
Moreover, the above characterization of the unimodular eigenvectors is independent of the strength of the disorder, i.e., the variance of the random variables $\theta_{x,j}^{(i)}$.
Any arbitrarily small, but finite disorder strength gives the same results and hence the above reasoning and the following arguments apply to almost clean and therefor almost translationally invariant systems.
In particular this means, that the disorder strength could be sent to zero after taking the thermodynamic limit first without altering the results.

The eigenvectors $\ket{r}$ constructed above are not orthogonal. In fact, noting that $|\eta_{2(r-s)}|= \gcd(|r-s|,t)$
the Gram matrix $G$ with
\begin{align}
G_{r,s} = \braket{r | s} = d^{-2(t - \gcd(|r-s|,t))}
\label{eq:gram_matrix_elements}
\end{align}
is not diagonal, see App.~\ref{app:weingarten_matrix} for details.
We hence write the projection onto the $t$-dimensional eigenspace corresponding to the leading eigenvalue 1 as
\begin{align}
    \Proj_t = \sum_{r,s=0}^{t-1}W_{r,s} \ket{r}\!\bra{s}.
\end{align}
Here $W$ is the so-called Weingarten matrix \cite{Col2003}
restricted to the cyclic subgroup of $S_{2t}$ generated by $\eta^2$.
It is the (pseudo-)inverse of the Gram matrix $G$.
Using  $\lim_{L\to \infty}\S_t^{L-l}=\Proj_t$ we obtain
\begin{align}
    K_A(t) = \lim_{L\to \infty}K_A(t; L) = \sum_{r,s=0}^{t-1}W_{r,s}\bra{s}\T_t^l\ket{r}
\end{align}
in the thermodynamic limit.
Defining the matrix $T$ by its matrix elements
\begin{align}
    T_{s,r} = \bra{s}\T_t^l\ket{r}
\end{align}
this can be compactly represented as
\begin{align} \label{eq:PSFF-viaTW}
    K_A(t) = \tr(T W).
\end{align}

That is, we have rephrased the computation of the PSFF into computing the trace of the product of two $t\times t$ matrices.
In the following we characterize these two matrices in more detail. Additionally we will see, that $T$ depends only on the dual-unitarity of the gates and is completely independent from the disorder realization.

\subsection{Partial Spectral Form Factor from Permutations}
\label{sec:psff_from_permutations}

We start with the computation of the Weingarten matrix $W$, which is equivalent to constructing an orthogonal basis of the eigenspace for eigenvalue $1$. This can be achieved systematically by utilizing the full symmetry of the transfer operator $\S_t$ under independent even site shifts in both the forward and backward time sheet \cite{GarCha2021,RidKeyProBer2024:p}.
For our purpose, it is, however, sufficient to
show that $W$ is close to the identity $\mathds{1}_t$. While a detailed proof of this claim is presented in App.~\ref{app:weingarten_matrix} this can be seen, by noting that
\begin{align}
   0 < G_{r,s} = d^{-2(t - \gcd(|r-s|,t))} \leq d^{-t},
   \label{eq:T_bound_matelements}
\end{align}
for $r \neq s$.

The above estimate implies that the Hilbert-Schmidt norm of $\mathds{1}_t - G$ is bounded as $\|\mathds{1}_t - G\|_{\mathrm{HS}} < td^{-t}$. As $G$ is close to the identity it is invertible and the inverse $W$ is close to the identity as well.
In fact, we obtain the bound
\begin{align}
\|\mathds{1}_t - W\|_{\mathrm{HS}} < ctd^{-t}
\label{eq:bound_W}
\end{align}
for some constant $c$ of order 1.
The interpretation of this is that the states $\ket{r}$ are almost orthogonal, i.e., up to some exponentially small overlapp $\sim d^{-t}$.
A notable special case in which $W$ can be obtained easily is for $t$ being a sufficiently large prime, see App.~\ref{app:weingarten_matrix}. In this case $G_{r,s}= d^{-2(t - 1)}$ for $r\neq s$ is much smaller than the bound~\eqref{eq:bound_G_elements}. In fact for arbitrary $t$ the bound corresponds to the worst case scenario, whereas typical matrix elements of $G$ are much stronger suppressed. Consequently, the bound on the norm of the Weingarten matrix $W$, Eq.~\eqref{eq:bound_W}, is not tight.\\

It remains to compute the  matrix elements of the $t\times t$ matrix $T$.
The main strategy is to use, that $\T_t^l$ essentially maps the periodic shifts $\ket{r}$ to permutations. Computing the overlap of the resulting permutation with $\ket{s}$ can be done solely on the level of permutations and their cycle decomposition yielding the matrix elements $T_{r,s}$.
This can be done analytically for the diagonal elements,
which are the only ones we need to evaluate
Eq.~\eqref{eq:PSFF-viaTW} as $W$ is close to the identity.

We start by simplifying $\T_t^l$ as
\begin{align}
\T_t^l = D_A \! \!\!\!\!\!
\begin{tikzpicture}[baseline=85, scale=0.7]
\foreach \x in {-1.1}{
    \foreach \y in {-0.8}{
        \draw[thick, Stealth - Stealth](\x, \y + 1.8) -- (\x, \y) -- (\x + 1.8, \y);
        \Text[x=\x + 1.95,y=\y -.25]{$x$}
        \Text[x=\x -0.32,y=\y + 1.8]{$t$}
}}
\foreach \x in {0,...,2}
{
    \foreach \y in {0,...,4}
    {
        \foldedgate{2*\x}{2*\y + 0.5}
    }
}
\foreach \x in {0,...,2}
{
    \foreach \y in {0,...,3}
    {
        \foldedgate{2*\x + 1}{2*\y + 1.5}
    }
}
\foreach \x in {1,...,5}{
    \draw[thick] (\x - 0.5, 0.) -- (\x - 0.5, -0.3);
    \draw[thick] (\x - 0.5, 9.) -- (\x - 0.5, 9.3);   
}
\draw[thick] (5.5, 9.3) -- (5.5, 9.1) -- (5.6, 9.) -- (5.8, 9.);
\draw[thick] (5.5, -0.3) -- (5.5, -0.1) -- (5.6, 0.) -- (5.8, 0.);
\foreach \y in {0,...,9}{
    \draw[thick] (-0.7, \y) -- (- 0.5, \y);
}
\foreach \y in {1,...,8}{
    \draw[thick] (5.5, \y) -- (5.8, \y);
}
\foreach \x in {1,...,6}{
    \IdState{\x-0.5}{9.3}
    \IdState{\x-0.5}{-0.3}
}
\end{tikzpicture} \, \,
= D_A \, \,   
\begin{tikzpicture}[baseline=85, scale=0.7]
\foreach \x in {0}
{
    \foreach \y in {0,...,4}
    {
        \foldedgate{2*\x}{2*\y + 0.5}
    }
}
\foreach \x in {0,2}
{
    \foreach \y in {0,...,3}
    {
        \foldedgate{2*\x + 1}{2*\y + 1.5}
    }
}
\foreach \x in {1,2}
{
    \foreach \y in {1,...,3}
    {
        \foldedgate{2*\x}{2*\y + 0.5}
    }
}
\foreach \x in {1}
{
    \foreach \y in {1,2}
    {
        \foldedgate{2*\x + 1}{2*\y + 1.5}
    }
}
\foreach \x in {1}{
    \draw[thick] (\x - 0.5, 0.) -- (\x - 0.5, -0.3);
    \draw[thick] (\x - 0.5, 9.) -- (\x - 0.5, 9.3);   
}
\draw[thick] (5.5, 9.3) -- (5.5, 9.1) -- (5.6, 9.) -- (5.8, 9.);
\draw[thick] (5.5, -0.3) -- (5.5, -0.1) -- (5.6, 0.) -- (5.8, 0.);
\foreach \y in {0,...,9}{
    \draw[thick] (-0.7, \y) -- (- 0.5, \y);
}
\foreach \y in {1,...,8}{
    \draw[thick] (5.5, \y) -- (5.8, \y);
}
\foreach \x in {1,6}{
    \IdState{\x-0.5}{9.3}
    \IdState{\x-0.5}{-0.3}
}
\foreach \x in {2,5}{
    \IdState{\x-0.5}{8.}
    \IdState{\x-0.5}{1.}
}
\foreach \x in {3,4}{
\IdState{\x-0.5}{7.}
\IdState{\x-0.5}{2.}
}
\end{tikzpicture}
\label{eq:transfer_operator_A_long_times}
\end{align}%
using the unitality of the folded gates, Eq.~\eqref{eq:dual_unitality}, depicted here for $t=5$ and $l=3$. \\

Rewriting this in the unfolded picture, the matrix elements $\bra{Y}\T_t^l\ket{X}= \tr\left(Y^\dagger \T_t^l\left(X\right)\right)$ for $X,Y \in B\left(\Hil_{2t}\right)$ (purple boxes below) can be diagrammatically represented as (rotated by $90^\circ$)
\begin{align}
\bra{Y}\T_t^l\ket{X}
& =  \begin{tikzpicture}[baseline=0, scale=0.7]
\foreach \x in {-2.8}{
    \foreach \y in {7.3}{
        \draw[thick, Stealth - Stealth](\x, \y - 1.8) -- (\x, \y) -- (\x + 1.8, \y);
        \Text[x=\x + 1.95,y=\y + .25]{$t$}
        \Text[x=\x -0.32,y=\y - 1.8]{$x$}
}}
\foreach \x in {0,...,4}
{
    \dualgate{2*\x}{1}
    \dualgateadjoint{2*\x}{- 1.}
}
\foreach \x in {0,...,3}
{
    \dualgate{2*\x + 1}{2}
    \dualgateadjoint{2*\x + 1}{- 2.}
    \dualgate{2*\x + 1}{6}
    \dualgateadjoint{2*\x + 1}{- 6.}
}
\foreach \x in {0,...,2}
{
    \dualgate{2*\x + 2}{3}
    \dualgateadjoint{2*\x + 2}{- 3.}
    \dualgate{2*\x + 2}{5}
    \dualgateadjoint{2*\x + 2}{- 5.}
}
\foreach \x in {0,...,1}
{
    \dualgate{2*\x + 3}{4}
    \dualgateadjoint{2*\x + 3}{- 4.}
}
\draw[thick, rounded corners=2pt] (- 0.5, 1.5) -- (-1., 1.5) -- (-1., -1.5) -- (-0.5, -1.5);
\draw[thick, rounded corners=2pt] (0.5, 2.5) -- (-1.25, 2.5) -- (-1.25, -2.5) -- (0.5, -2.5);
\draw[thick, rounded corners=2pt] (1.5, 3.5) -- (-1.5, 3.5) -- (-1.5, -3.5) -- (1.5, -3.5);
\draw[thick, rounded corners=2pt] (1.5, 4.5) -- (-1.75, 4.5) -- (-1.75, -4.5) -- (1.5, -4.5);
\draw[thick, rounded corners=2pt] (0.5, 5.5) -- (-2., 5.5) -- (-2., -5.5) -- (0.5, -5.5);
\draw[thick, rounded corners=2pt] (-0.5, 6.8) -- (-0.5, 6.5) -- (-2.25, 6.5) -- (-2.25, -6.5) -- (-0.5, -6.5) -- (-0.5, -6.8);
\draw[thick, rounded corners=2pt] (8.5, 1.5) -- (9., 1.5) -- (9., -1.5) -- (8.5, -1.5);
\draw[thick, rounded corners=2pt] (7.5, 2.5) -- (9.25, 2.5) -- (9.25, -2.5) -- (7.5, -2.5);
\draw[thick, rounded corners=2pt] (6.5, 3.5) -- (9.5, 3.5) -- (9.5, -3.5) -- (6.5, -3.5);
\draw[thick, rounded corners=2pt] (6.5, 4.5) -- (9.75, 4.5) -- (9.75, -4.5) -- (6.5, -4.5);
\draw[thick, rounded corners=2pt] (7.5, 5.5) -- (10., 5.5) -- (10., -5.5) -- (7.5, -5.5);
\draw[thick, rounded corners=2pt] (8.5, 6.8) -- (8.5, 6.5) -- (10.25, 6.5) -- (10.25, -6.5) -- (8.5, -6.5) -- (8.5, -6.8);
\draw[thick, color=black, fill=myrose, rounded corners=2pt]  (-0.85, -0.4) rectangle  (8.85, 0.4);
\draw[thick, color=black, fill=myrose, rounded corners=2pt]  (-0.85, -6.8) rectangle  (8.85, -7.6);
\foreach \x in {0,...,9}{
    \draw[thick] (\x - 0.5, 0.4) -- (\x - 0.5,  0.5);
    \draw[thick] (\x - 0.5, -0.4) -- (\x - 0.5,  -0.5);
    \draw[thick] (\x - 0.5, -7.6) -- (\x - 0.5,  -7.9);
}
\foreach \x in {1,...,8}{
    \draw[thick] (\x - 0.5, 6.5) -- (\x - 0.5,  6.8);
    \draw[thick] (\x - 0.5, -6.5) -- (\x - 0.5,  -6.8);
}
\foreach \x in {0,...,9}{
    \draw[thick] (\x-0.3, 6.8 ) arc (0:180: 0.1);
    \draw[thick] (\x - 0.5, -7.9 ) arc (0:180: -0.1);
    \foreach \y in {1,2,3}{
        \draw[thick](\x - 0.3, 6.85 - 0.1*\y) -- (\x - 0.3,  6.8 - 0.1*\y);
        \draw[thick](\x - 0.3, -7.95 + 0.1*\y) -- (\x - 0.3,  -7.9 + 0.1*\y);
    }
}
\draw (4.,0) node {$X$};
\draw (4.,-7.2) node {$Y^\dagger$};
\end{tikzpicture} \, .
\end{align}%
Using the cyclic property of the trace, this can be cast in the more symmetric form
\begin{align}
\bra{Y}\T_t^l\ket{X}
& = \, \,  \begin{tikzpicture}[baseline=-58, scale=0.7]
\foreach \x in {0,...,4}
{
    \dualgate{2*\x}{1}
    \dualgateadjoint{2*\x}{- 1.}
}
\foreach \x in {0,...,3}
{
    \dualgate{2*\x + 1}{2}
    \dualgateadjoint{2*\x + 1}{- 2.}
}
\foreach \x in {0,...,2}
{
    \dualgate{2*\x + 2}{3}
    \dualgateadjoint{2*\x + 2}{- 3.}
}
\foreach \x in {0,...,1}
{
    \dualgate{2*\x + 3}{4}
    \dualgateadjoint{2*\x + 3}{- 4.}
}
\draw[thick, rounded corners=2pt] (- 0.5, 1.5) -- (-1., 1.5) -- (-1., -1.5) -- (-0.5, -1.5);
\draw[thick, rounded corners=2pt] (0.5, 2.5) -- (-1.25, 2.5) -- (-1.25, -2.5) -- (0.5, -2.5);
\draw[thick, rounded corners=2pt] (1.5, 3.5) -- (-1.5, 3.5) -- (-1.5, -3.5) -- (1.5, -3.5);
\draw[thick, rounded corners=2pt] (8.5, 1.5) -- (9., 1.5) -- (9., -1.5) -- (8.5, -1.5);
\draw[thick, rounded corners=2pt] (7.5, 2.5) -- (9.25, 2.5) -- (9.25, -2.5) -- (7.5, -2.5);
\draw[thick, rounded corners=2pt] (6.5, 3.5) -- (9.5, 3.5) -- (9.5, -3.5) -- (6.5, -3.5);
\draw[thick, color=black, fill=myrose, rounded corners=2pt]  (-0.85, -0.4) rectangle  (8.85, 0.4);
\foreach \x in {0,...,9}{
    \draw[thick] (\x - 0.5, 0.4) -- (\x - 0.5,  0.5);
    \draw[thick] (\x - 0.5, -0.4) -- (\x - 0.5,  -0.5);
}
\draw (4.,0) node {$X$};
\foreach \x in {0,...,2}
{
    \dualgateadjoint{2*\x + 2}{-5}
    \dualgate{2*\x + 2}{- 9.}
}
\foreach \x in {0,...,3}
{
    \dualgateadjoint{2*\x + 1}{-6}
    \dualgate{2*\x + 1}{- 8.}
}
\draw[thick, rounded corners=2pt] (1.5, -4.5) -- (-1.5, -4.5) -- (-1.5, -9.5) -- (1.5, -9.5);
\draw[thick, rounded corners=2pt] (0.5, -5.5) -- (-1.25, -5.5) -- (-1.25, -8.5) -- (.5, -8.5);
\draw[thick, rounded corners=2pt] (-0.5, -6.6) -- (-0.5, -6.3) -- (-1., -6.3) -- (-1., -7.7) -- (-.5, -7.7) -- (-.5, -7.4);
\draw[thick, rounded corners=2pt] (6.5, -4.5) -- (9.5, -4.5) -- (9.5, -9.5) -- (6.5, -9.5);
\draw[thick, rounded corners=2pt] (7.5, -5.5) -- (9.25, -5.5) -- (9.25, -8.5) -- (7.5, -8.5);
\draw[thick, rounded corners=2pt] (8.5, -6.6) -- (8.5, -6.3) -- (9., -6.3) -- (9., -7.7) -- (8.5, -7.7) -- (8.5, -7.4);
\draw[thick, color=black, fill=myrose, rounded corners=2pt]  (-0.85, -7.4) rectangle  (8.85, -6.6);
\foreach \x in {1,...,8}{
    \draw[thick] (\x - 0.5, -7.4) -- (\x - 0.5,  -7.5);
    \draw[thick] (\x - 0.5, -6.6) -- (\x - 0.5,  -6.5);
}
\draw (4.,-7.) node {$Y^\dagger$};
\foreach \x in {3,...,6}{
    \draw[thick](\x - 0.5, 4.5) -- (\x - 0.5, 4.8);
    \draw[thick](\x - 0.5, -9.5) -- (\x - 0.5, -9.8);
    \draw[thick] (\x-0.3, 4.8 ) arc (0:180: 0.1);
    \draw[thick] (\x - 0.5, -9.8 ) arc (0:180: -0.1);
    \foreach \y in {1,2,3}{
        \draw[thick](\x - 0.3, 4.85 - 0.1*\y) -- (\x - 0.3,  4.8 - 0.1*\y);
        \draw[thick](\x - 0.3, -9.85 + 0.1*\y) -- (\x - 0.3,  -9.8 + 0.1*\y);
    }
}
\end{tikzpicture} \,  .
\label{eq:transfer_operator_T_matrix_elements_channel}
\end{align}%
In the following we aim for contracting this tensor network exactly for $X=\mathds{P}_{\eta^{2r}}/d^{t}$ and $Y=\mathds{P}_{\eta^{-2s}}/d^{t}$, mostly focusing on $r=s$.
This will be accomplished by first proving that in this case all nontrivial 
gates can be replaced with the respective identity. To see this, we first 
introduce some notation. For $0\leq k \leq l$ we define the set of lattice sites 
$M_{t, k}=\{1,2,\ldots, k\}\cup \{2t-k+1,2t-k+2,\ldots,2t\}$ and its complement 
$K_{t,k}=\{k+1, k+2, \ldots, 2t - k\}$ such that $L_t = M_{t,k} \cup K_{t,k}$.
We find, that taking the partial trace over the degrees of freedom in 
$M_{t,k}$ of a permutation $\pi \in S_{L_t}$ acting on the full lattice $L_t$ 
yields a permutation $\pi^\prime \in S_{K_{t,k}}$ acting on the lattice sites 
in $K_{t,k}$ up to a factor. More precisely, on has
\begin{align}
\tr_{M_{t,k}}\left(\Perm_{\pi}\right) = d^{|\pi|_{M_{t,k}}}\Perm_{\pi^\prime}
\label{eq:partial_trace_permutation}
\end{align}
with $|\pi|_{M_{t,k}}$ denoting the number of cycles of $\pi$ completely contained in $M_{t,k}$.
The permutation $\pi^\prime$ is determined by $\pi^{\prime}(x) = \pi^{\tau(\pi, x)}(x)$ for $x \in K_{t,k}$ and with $\tau(\pi, x)$ being the smallest non-negative integer $\tau^\prime$ such that $\pi^{\tau^\prime}(x) \in K_{t,k}$.
We provide details on this construction in terms of the induced map  $S_{L_t} \to S_{K_{t,k}}, \pi \mapsto \pi^\prime$ in App.~\ref{app:proofs} and illustrate it here with an example only.

For instance, consider the permutation \begin{align}
    \pi = \left(\begin{matrix}
        1 & 2 & 3 & 4 & 5 & 6 & 7 & 8 & 9 & 10 \\
        2 & 1 & 5 & 3 & 7 & 8 & 4 & 10 & 9 & 6
    \end{matrix}\right) \quad \text{leading to} \quad
     \pi^\prime = \left(\begin{matrix}
        4 & 5 & 6 & 7 \\
        5 & 7 & 6 & 4
    \end{matrix}\right)
\end{align}
which can be verified diagrammatically as
\begin{align}
\tr_{M_{t,k}}\left(\mathds{P}_\pi\right) = \, \,
\begin{tikzpicture}[baseline=-2, scale=0.7]
\draw[thick, rounded corners=2pt] (- 0.5, 1. )-- (- 0.5, 1.5 ) -- (-1., 1.5 ) -- (-1., -1.5 ) -- (-0.5, -1.5 ) -- (-0.5, -1. );
\draw[thick, rounded corners=2pt] (0.5, 1.)-- (0.5,1.75) -- (-1.25, 1.75) -- (-1.25, -1.75 ) -- (0.5, -1.75 ) -- (0.5, -1. );
\draw[thick, rounded corners=2pt] (1.5, 1. )-- (1.5,2. ) -- (-1.5, 2. ) -- (-1.5, -2. ) -- (1.5, -2. ) -- (1.5, -1. );
\draw[thick, rounded corners=2pt] (8.5, 1. ) -- (8.5, 1.5 ) -- (9., 1.5 ) -- (9., -1.5 ) -- (8.5, -1.5 ) -- (8.5, -1. );
\draw[thick, rounded corners=2pt] (7.5, 1. ) -- (7.5, 1.75 ) -- (9.25, 1.75 ) -- (9.25, -1.75 ) -- (7.5, -1.75 ) -- (7.5, -1. );
\draw[thick, rounded corners=2pt] (6.5, 1. ) -- (6.5, 2. ) -- (9.5, 2. ) -- (9.5, -2. ) -- (6.5, -2. ) -- (6.5, -1. );
\foreach \x in {3, 4, 5, 6}{
\draw[thick, rounded corners=2pt] (\x - 0.5, 1.) -- ( \x - 0.5, 2.3);
\draw[thick, rounded corners=2pt] (\x - 0.5, -1.) -- ( \x - 0.5, -2.3);
}
\draw[ultra thick, rounded corners=2pt, color=red] ( - 0.5, -1.) -- ( - 0.5, -0.7) -- (0.5, 0.7) --(0.5, 1.);
\draw[ultra thick, rounded corners=2pt, color=red] ( - 0.5, 1.) -- ( - 0.5, 0.7) -- (0.5, -0.7) --(0.5, -1.);
\draw[ultra thick, rounded corners=2pt, color=myorange] ( 1.5, -1.) -- ( 1.5, -0.7) -- (3.5, 0.7) --(3.5, 1.);
\draw[ultra thick, rounded corners=2pt, color=myorange] ( 2.5, -1.) -- ( 2.5, -0.7) -- (1.5, 0.7) --(1.5, 1.);
\draw[ultra thick, rounded corners=2pt, color=mygreen] ( 3.5, -1.) -- ( 3.5, -0.7) -- (5.5, 0.7) --(5.5, 1.);
\draw[ultra thick, rounded corners=2pt, color=myorange] ( 4.5, -1.) -- ( 4.5, -0.7) -- (6.5, 0.7) --(6.5, 1.);
\draw[ultra thick, rounded corners=2pt, color=mygreen] ( 5.5, -1.) -- ( 5.5, -0.7) -- (2.5, 0.7) --(2.5, 1.);
\draw[ultra thick, rounded corners=2pt, color=myorange] ( 6.5, -1.) -- ( 6.5, -0.7) -- (8.5, 0.7) --(8.5, 1.);
\draw[ultra thick, rounded corners=2pt, color=red] ( 7.5, -1.) -- ( 7.5, -0.7) -- (7.5, 0.7) --(7.5, 1.);
\draw[ultra thick, rounded corners=2pt, color=myorange] ( 8.5, -1.) -- ( 8.5, -0.7) -- (4.5, 0.7) --(4.5, 1.);
\end{tikzpicture} \, \, = d^2 \, \,
\begin{tikzpicture}[baseline=-2, scale=0.7]
\draw[ultra thick, rounded corners=2pt, color=myorange] ( 2.5, -1.) -- ( 2.5, -0.7) -- (3.5, 0.7) --(3.5, 1.);
\draw[ultra thick, rounded corners=2pt, color=mygreen] ( 3.5, -1.) -- ( 3.5, -0.7) -- (5.5, 0.7) --(5.5, 1.);
\draw[ultra thick, rounded corners=2pt, color=myorange] ( 4.5, -1.) -- ( 4.5, -0.7) -- (4.5, 0.7) --(4.5, 1.);
\draw[ultra thick, rounded corners=2pt, color=mygreen] ( 5.5, -1.) -- ( 5.5, -0.7) -- (2.5, 0.7) --(2.5, 1.);
\foreach \x in {3, 4, 5, 6}{
    \draw[thick, rounded corners=2pt] (\x - 0.5, 1.) -- ( \x - 0.5, 1.5);
    \draw[thick, rounded corners=2pt] (\x - 0.5, -1.) -- ( \x - 0.5, -1.5);
}
\end{tikzpicture}
\, \, = d^2 \mathds{P}_{\pi^\prime}. \label{eq:permuations_example}
\end{align}%
Each cycle of $\pi$ that is completely contained in $M_{t,k}$ gives a factor of $\tr(\mathds{1}_d)=d$. In the above example, there are two such cycles,  $(1,2)$ and the fixed point $9$, i.e. a cycle with just one element,
 indicated by the red wires, yielding $|\pi|_{M_{t,k}}=2$.
In contrast, the lattice sites $5,7 \in K_{t,k}$ are mapped to sites in $K_{t,k}$ by $\pi$, corresponding to $\tau(\pi,x)=1$ and are thus not affected by the partial trace over $M_{t,k}$. They are depicted by the green wires above.
The final case is that of the sites $4,6 \in K_{t,k}$, which are mapped to sites in $M_{t,k}$ by $\pi$, indicated by orange wires. They "loop" around in $M_{t,k}$ until they are finally mapped to some point in $K_{t,l}$ corresponding to $\tau(\pi,4)=2$ and $\tau(\pi,6)=3$, respectively.  \\

For $\pi=\eta^{2r}$ the resulting operators $\tr_{M_{t,k}}\left(\Perm_{\eta^{2r}}\right)$ are invariant under conjugation by $\tilde{U}_0^{\otimes t-k}$, i.e, they obey
\begin{align}
\tilde{U}_0^{\otimes t-k}\,\tr_{M_{t,k}}\!\left(\mathds{P}_{\eta^{2r}}\right)\left(\tilde{U}_0^\dagger\right)^{\otimes t-k} & = \, \, \begin{tikzpicture}[baseline=-2, scale=0.7]
\foreach \x in {1,...,3}
{
    \dualgate{2*\x}{1}
    \dualgateadjoint{2*\x}{- 1.}
}
\draw[thick, color=black, fill=myrose, rounded corners=2pt]  (-0.85, -0.4) rectangle  (8.85, 0.4);
\draw[thick, rounded corners=2pt] (- 0.5, 0.4)-- (- 0.5, 0.75) -- (-1., 0.75) -- (-1., -0.75) -- (-0.5, -0.75) -- (-0.5, -0.4);
\draw[thick, rounded corners=2pt] (0.5, 0.4)-- (0.5,1.25) -- (-1.25, 1.25) -- (-1.25, -1.25) -- (0.5, -1.25) -- (0.5, -0.4);
\draw[thick, rounded corners=2pt] (8.5, 0.4) -- (8.5, 0.75) -- (9., 0.75) -- (9., -0.75) -- (8.5, -0.75) -- (8.5, -0.4);
\draw[thick, rounded corners=2pt] (7.5, 0.4) -- (7.5, 1.25) -- (9.25, 1.25) -- (9.25, -1.25) -- (7.5, -1.25) -- (7.5, -0.4);
\draw (4.,0) node {$\eta^{2r}$};
\foreach \x in {1,...,6}{
\draw[thick, rounded corners=2pt] (\x + 0.5, 0.4)-- (\x +  0.5, 0.5);
\draw[thick, rounded corners=2pt] (\x + 0.5, 1.5)-- (\x +  0.5, 1.6);
\draw[thick, rounded corners=2pt] (\x + 0.5, -0.4)-- (\x +  0.5, -0.5);
\draw[thick, rounded corners=2pt] (\x + 0.5, -1.5)-- (\x +  0.5, -1.6);
}
\end{tikzpicture} \label{eq:iterative_cancelation0} \\
& = \, \, \begin{tikzpicture}[baseline=-2, scale=0.7]
\draw[thick, color=black, fill=myrose, rounded corners=2pt]  (-0.85, -0.4) rectangle  (8.85, 0.4);
\draw[thick, rounded corners=2pt] (- 0.5, 0.4)-- (- 0.5, 0.75) -- (-1., 0.75) -- (-1., -0.75) -- (-0.5, -0.75) -- (-0.5, -0.4);
\draw[thick, rounded corners=2pt] (0.5, 0.4)-- (0.5,1.25) -- (-1.25, 1.25) -- (-1.25, -1.25) -- (0.5, -1.25) -- (0.5, -0.4);
\draw[thick, rounded corners=2pt] (8.5, 0.4) -- (8.5, 0.75) -- (9., 0.75) -- (9., -0.75) -- (8.5, -0.75) -- (8.5, -0.4);
\draw[thick, rounded corners=2pt] (7.5, 0.4) -- (7.5, 1.25) -- (9.25, 1.25) -- (9.25, -1.25) -- (7.5, -1.25) -- (7.5, -0.4);
\draw (4.,0) node {$\eta^{2r}$};
\foreach \x in {1,...,6}{
    \draw[thick, rounded corners=2pt] (\x + 0.5, 0.4)-- (\x +  0.5, 1.5);
    \draw[thick, rounded corners=2pt] (\x + 0.5, -0.4)-- (\x +  0.5, -1.5);
}
\end{tikzpicture} \\
&  = \tr_{M_{t,k}}\!\left(\mathds{P}_{\eta^{2r}}\right)
\label{eq:iterative_cancelation}
\end{align}%
for any $k$.
This is a consequence of dual unitarity of the gates, Eq.~\eqref{eq:dual_unitarity}.
As an example we illustrate the simple case $k=2$, $r=1$, and $t=5$ diagrammatically as
\begin{align}
\tilde{U}_0^{\otimes t - k}\,\tr_{M_{t,k}}\left(\mathds{P}_{\eta^{2r}}\right)\left(\tilde{U}_0^\dagger\right)^{\otimes t - k} & = \, \,
\begin{tikzpicture}[baseline=-9, scale=0.7]
\foreach \x in {1,...,3}
{
    \dualgateadjoint{2*\x.}{-1.5}
    \dualgate{2*\x}{1.}
}
 \draw[thick, rounded corners=2pt] (-0.5, 0.5) -- (-0.5, 0.75) -- (-2., 0.75) --(-2., -1.25) -- (-0.5, -1.25) -- (-0.5, -1.);
\draw[thick, rounded corners=2pt] (0.5, 0.5) -- (0.5, 1.) -- (-2.25, 1.) --(-2.25, -1.5) -- (0.5, -1.5) -- (0.5, -1.);
\draw[thick, rounded corners=2pt] (8.5, 0.5) -- (8.5, 0.75) -- (10., 0.75) --(10., -1.25) -- (8.5, -1.25) -- (8.5, -1.);
\draw[thick, rounded corners=2pt] (7.5, 0.5) -- (7.5, 1.) -- (10.25, 1.) --(10.25, -1.5) -- (7.5, -1.5) -- (7.5, -1.);
\foreach \x in {0,...,7}{
    \draw[thick, rounded corners=2pt] (\x - 0.5, -1.) -- (\x - 0.5, -0.75)  -- (\x + 2 - 0.5, 0.25) --  (\x + 2 - 0.5, 0.5);
}
\draw[thick, rounded corners=2pt] (8 - 0.5, -1.) -- (8 - 0.5, -0.75)  -- (8 + 1.5 - 0.5, 0.)  -- (8 + 1.5 - 0.2, 0.);
\draw[thick, rounded corners=2pt] (9 - 0.5, -1.) -- (9 - 0.5, -0.75)  -- (9 + .5 - 0.5, -0.5)  -- (9 + .5 - 0.2, -0.5);
\draw[thick, rounded corners=2pt] (- 0.5, 0.5) -- (- 0.5, 0.25)  -- (-1., 0.)  -- (-1.3, 0.);
\draw[thick, rounded corners=2pt] (0.5, 0.5) -- (0.5, 0.25)  -- (-1., -0.5)  -- (-1.3, -0.5);
\draw[thick] (-1.3, -0.7) arc (270:90: 0.1);
\draw[thick] (-1.3, -0.) arc (270:90: 0.1);
\draw[thick] (9.3, -0.7) arc (90:270: -0.1);
\draw[thick] (9.3, -0.) arc (90:270: -0.1);
\foreach \x in {1,2,3}{
    \draw[thick](-1.3 + 0.1*\x - 0.05, -0.7) -- (-1.3 + 0.1*\x + 0., -0.7);
    \draw[thick](-1.3 + 0.1*\x - 0.05, 0.2) -- (-1.3 + 0.1*\x + 0., 0.2);
    \draw[thick](9.3 - 0.1*\x + 0.05, -0.7) -- (9.3 - 0.1*\x + 0., -0.7);
    \draw[thick](9.3 - 0.1*\x + 0.05, 0.2) -- (9.3 - 0.1*\x + 0., 0.2);
}
\end{tikzpicture} \\
& = \, \,
\begin{tikzpicture}[baseline=-9, scale=0.7]
\foreach \x in {1,...,2}
{
    \dualgateadjoint{2*\x.}{-1.5}
    \dualgate{2*\x + 2.}{1.}
}
\draw[thick, rounded corners=2pt] (1.5, 0.5) -- (1.7, 0.7) -- (1.7, 1.3) --(1.5, 1.5);
\draw[thick, rounded corners=2pt] (2.5, 0.5) -- (2.3, 0.7) -- (2.3, 1.3) --(2.5, 1.5);
\draw[thick, rounded corners=2pt] (5.5, -1.) -- (5.7, -1.2) -- (5.7, -1.8) --(5.5, -2.);
\draw[thick, rounded corners=2pt] (6.5, -1.) -- (6.3, -1.2) -- (6.3, -1.8) --(6.5, -2.);
\draw[thick, rounded corners=2pt] (-0.5, 0.5) -- (-0.5, 0.75) -- (-2., 0.75) --(-2., -1.25) -- (-0.5, -1.25) -- (-0.5, -1.);
\draw[thick, rounded corners=2pt] (0.5, 0.5) -- (0.5, 1.) -- (-2.25, 1.) --(-2.25, -1.5) -- (0.5, -1.5) -- (0.5, -1.);
\draw[thick, rounded corners=2pt] (8.5, 0.5) -- (8.5, 0.75) -- (10., 0.75) --(10., -1.25) -- (8.5, -1.25) -- (8.5, -1.);
\draw[thick, rounded corners=2pt] (7.5, 0.5) -- (7.5, 1.) -- (10.25, 1.) --(10.25, -1.5) -- (7.5, -1.5) -- (7.5, -1.);
\foreach \x in {0,...,7}{
    \draw[thick, rounded corners=2pt] (\x - 0.5, -1.) -- (\x - 0.5, -0.75)  -- (\x + 2 - 0.5, 0.25) --  (\x + 2 - 0.5, 0.5);
}
\draw[thick, rounded corners=2pt] (8 - 0.5, -1.) -- (8 - 0.5, -0.75)  -- (8 + 1.5 - 0.5, 0.)  -- (8 + 1.5 - 0.2, 0.);
\draw[thick, rounded corners=2pt] (9 - 0.5, -1.) -- (9 - 0.5, -0.75)  -- (9 + .5 - 0.5, -0.5)  -- (9 + .5 - 0.2, -0.5);
\draw[thick, rounded corners=2pt] (- 0.5, 0.5) -- (- 0.5, 0.25)  -- (-1., 0.)  -- (-1.3, 0.);
\draw[thick, rounded corners=2pt] (0.5, 0.5) -- (0.5, 0.25)  -- (-1., -0.5)  -- (-1.3, -0.5);
\draw[thick] (-1.3, -0.7) arc (270:90: 0.1);
\draw[thick] (-1.3, -0.) arc (270:90: 0.1);
\draw[thick] (9.3, -0.7) arc (90:270: -0.1);
\draw[thick] (9.3, -0.) arc (90:270: -0.1);
\foreach \x in {1,2,3}{
    \draw[thick](-1.3 + 0.1*\x - 0.05, -0.7) -- (-1.3 + 0.1*\x + 0., -0.7);
    \draw[thick](-1.3 + 0.1*\x - 0.05, 0.2) -- (-1.3 + 0.1*\x + 0., 0.2);
    \draw[thick](9.3 - 0.1*\x + 0.05, -0.7) -- (9.3 - 0.1*\x + 0., -0.7);
    \draw[thick](9.3 - 0.1*\x + 0.05, 0.2) -- (9.3 - 0.1*\x + 0., 0.2);
}
\end{tikzpicture} \\
& = \ldots \\
& = \, \,
\begin{tikzpicture}[baseline=-9, scale=0.7]
\foreach \x in {0,2,4}{
    \draw[thick, rounded corners=2pt] (\x + 1.5, 0.5) -- (\x + 1.7, 0.7) -- (\x + 1.7, 1.3) --(\x + 1.5, 1.5);
    \draw[thick, rounded corners=2pt] (\x + 2.5, 0.5) -- (\x + 2.3, 0.7) -- (\x +  2.3, 1.3) --(\x + 2.5, 1.5);
    \draw[thick, rounded corners=2pt] (5.5 - \x, -1.) -- (5.7 - \x, -1.2) -- (5.7 - \x, -1.8) --(5.5 - \x, -2.);
    \draw[thick, rounded corners=2pt] (6.5 - \x, -1.) -- (6.3 - \x, -1.2) -- (6.3 - \x, -1.8) --(6.5 - \x, -2.);
}
\draw[thick, rounded corners=2pt] (-0.5, 0.5) -- (-0.5, 0.75) -- (-2., 0.75) --(-2., -1.25) -- (-0.5, -1.25) -- (-0.5, -1.);
\draw[thick, rounded corners=2pt] (0.5, 0.5) -- (0.5, 1.) -- (-2.25, 1.) --(-2.25, -1.5) -- (0.5, -1.5) -- (0.5, -1.);
\draw[thick, rounded corners=2pt] (8.5, 0.5) -- (8.5, 0.75) -- (10., 0.75) --(10., -1.25) -- (8.5, -1.25) -- (8.5, -1.);
\draw[thick, rounded corners=2pt] (7.5, 0.5) -- (7.5, 1.) -- (10.25, 1.) --(10.25, -1.5) -- (7.5, -1.5) -- (7.5, -1.);
\foreach \x in {0,...,7}{
    \draw[thick, rounded corners=2pt] (\x - 0.5, -1.) -- (\x - 0.5, -0.75)  -- (\x + 2 - 0.5, 0.25) --  (\x + 2 - 0.5, 0.5);
}
\draw[thick, rounded corners=2pt] (8 - 0.5, -1.) -- (8 - 0.5, -0.75)  -- (8 + 1.5 - 0.5, 0.)  -- (8 + 1.5 - 0.2, 0.);
\draw[thick, rounded corners=2pt] (9 - 0.5, -1.) -- (9 - 0.5, -0.75)  -- (9 + .5 - 0.5, -0.5)  -- (9 + .5 - 0.2, -0.5);
\draw[thick, rounded corners=2pt] (- 0.5, 0.5) -- (- 0.5, 0.25)  -- (-1., 0.)  -- (-1.3, 0.);
\draw[thick, rounded corners=2pt] (0.5, 0.5) -- (0.5, 0.25)  -- (-1., -0.5)  -- (-1.3, -0.5);
\draw[thick] (-1.3, -0.7) arc (270:90: 0.1);
\draw[thick] (-1.3, -0.) arc (270:90: 0.1);
\draw[thick] (9.3, -0.7) arc (90:270: -0.1);
\draw[thick] (9.3, -0.) arc (90:270: -0.1);
\foreach \x in {1,2,3}{
    \draw[thick](-1.3 + 0.1*\x - 0.05, -0.7) -- (-1.3 + 0.1*\x + 0., -0.7);
    \draw[thick](-1.3 + 0.1*\x - 0.05, 0.2) -- (-1.3 + 0.1*\x + 0., 0.2);
    \draw[thick](9.3 - 0.1*\x + 0.05, -0.7) -- (9.3 - 0.1*\x + 0., -0.7);
    \draw[thick](9.3 - 0.1*\x + 0.05, 0.2) -- (9.3 - 0.1*\x + 0., 0.2);
}
\end{tikzpicture} \\
& = \tr_{M_{t,k}}\left(\mathds{P}_{\eta^{2r}}\right).
\label{eq:cancelations_gates_shifts}
\end{align}%
Here, we first use dual-unitarity for the rightmost gate $\tilde{U}_0$ and the leftmost gate $\tilde{U}_0^\dagger$ by following the corresponding wires through the network.
Subsequently, we similarly use dual unitarity for the remaining two pairs of gates.
The general case for arbitrary $k$ is proven in App.~\ref{app:proofs} and is essentially a consequence of the fact that taking the partial trace over $M_{t,k}$ is compatible with the even-odd structure of the lattice $K_{t,k}$ and the unitaries $\tilde{U}_0^{\otimes t-k}$.
An analogous property holds with the roles of $\tilde{U}_0$ and its Hermitian conjugate exchanged.
We emphasize that even in the presence of disorder, we would only pair the space-time dual of the disordered gates with their respective adjoint. Hence the above cancellation still occurs. It is ultimately a consequence of unitarity and dual unitarity alone. Consequently, the matrix $T$ and its matrix elements  $T_{r,s} = \bra{s}\T_t^l\ket{r}$ are completely independent of the local gate and the disorder. This again justifies to neglect the disorder and the corresponding average in subsystem $A$ completely, as done in Eq.~\eqref{eq:psff_transferops_tdl}. \\

Applying the above properties, Eqs.~\eqref{eq:iterative_cancelation0}-\eqref{eq:iterative_cancelation}, for both $\mathds{P}_{\eta^{2r}}$ and $\mathds{P}_{\eta^{-2s}}$, iteratively for $k$ from $0$ to at most $l$ all the non-trivial gates in Eq.~\eqref{eq:transfer_operator_T_matrix_elements_channel} cancel.
We thus eventually arrive at the simplified expression for the matrix elements given by
\begin{align}
T_{r,s} & = d^{-2t} \, \, \begin{tikzpicture}[baseline=-30, scale=0.7]
\foreach \y in {0,-3}{
\draw[thick, color=black, fill=myrose, rounded corners=2pt]  (-0.85, -0.4  +\y) rectangle  (8.85, 0.4 +\y);
\draw[thick, rounded corners=2pt] (- 0.5, 0.4 +\y)-- (- 0.5, 0.75 +\y) -- (-1., 0.75 +\y) -- (-1., -0.75 +\y) -- (-0.5, -0.75 +\y) -- (-0.5, -0.4 +\y);
\draw[thick, rounded corners=2pt] (0.5, 0.4 +\y)-- (0.5,1. +\y) -- (-1.25, 1. +\y) -- (-1.25, -1. +\y) -- (0.5, -1. +\y) -- (0.5, -0.4 +\y);
\draw[thick, rounded corners=2pt] (1.5, 0.4 +\y)-- (1.5,1.25 +\y) -- (-1.5, 1.25 +\y) -- (-1.5, -1.25 +\y) -- (1.5, -1.25 +\y) -- (1.5, -0.4 +\y);
\draw[thick, rounded corners=2pt] (8.5, 0.4 +\y) -- (8.5, 0.75 +\y) -- (9., 0.75 +\y) -- (9., -0.75 +\y) -- (8.5, -0.75 +\y) -- (8.5, -0.4 +\y);
\draw[thick, rounded corners=2pt] (7.5, 0.4 +\y) -- (7.5, 1. +\y) -- (9.25, 1. +\y) -- (9.25, -1. +\y) -- (7.5, -1. +\y) -- (7.5, -0.4 +\y);
\draw[thick, rounded corners=2pt] (6.5, 0.4 +\y) -- (6.5, 1.25 +\y) -- (9.5, 1.25 +\y) -- (9.5, -1.25 +\y) -- (6.5, -1.25 +\y) -- (6.5, -0.4 +\y);
\foreach \x in {2,...,5}{
    \draw[thick, rounded corners=2pt] (\x + 0.5, 0.4 +\y)-- (\x +  0.5, 1.5 +\y);
    \draw[thick, rounded corners=2pt] (\x + 0.5, -0.4 +\y)-- (\x +  0.5, -1.5 +\y);
}
}
\draw (4.,0) node {$\eta^{2r}$};
\draw (4.,-3) node {$\eta^{-2s}$};
\foreach \x in {3,...,6}{
    \draw[thick] (\x-0.3, 1.5 ) arc (0:180: 0.1);
    \draw[thick] (\x - 0.5, -4.5 ) arc (0:180: -0.1);
    \foreach \y in {1,2,3}{
        \draw[thick](\x - 0.3, 1.55 - 0.1*\y) -- (\x - 0.3,  1.5 - 0.1*\y);
        \draw[thick](\x - 0.3, -4.55 + 0.1*\y) -- (\x - 0.3,  -4.5 + 0.1*\y);
    }
}
\end{tikzpicture} \\
  &  = d^{-2t + |\eta^{2r}|_{M_{t,l}} + |\eta^{-2s}|_{M_{t,l}}} \tr_{K_{t,l}}\left(\mathds{P}_{(\eta^{-2s})^\prime }\mathds{P}_{(\eta^{2r})^\prime }\right) \label{eq:matrix_elements_from_permuations1} \\ 
& = d^{-2t + |\eta^{2r}|_{M_{t,l}} + |\eta^{2s}|_{M_{t,l}}} \tr_{K_{t,l}}\left(\mathds{P}_{(\eta^{-2s})^\prime (\eta^{2r})^\prime }\right)  \label{eq:matrix_elements_from_permuations2} \\
& = d^{-2t + |\eta^{2r}|_{M_{t,l}} + |\eta^{2s}|_{M_{t,l}} + |(\eta^{-2s})^\prime (\eta^{2r})^\prime |}\, .
\label{eq:matrix_elements_from_permuations}
\end{align}%
From line~\eqref{eq:matrix_elements_from_permuations1}~to~\eqref{eq:matrix_elements_from_permuations2}
we use $|\pi^{-1}|_{M_{t,l}}=|\pi|_{M_{t,l}}$ and the fact that $\mathds{P}$ is a representation, i.e., a group homomorphism.
The above expression can be evaluated by numerically counting the numbers of cycles in the respective permuations. This is used in Sec.~\ref{sec:numerics} to obtain numerical results directly in the thermodynamic limit.
For our purpose it is, however, sufficient to compute the diagonal matrix elements only.
To do so, we note that from the diagrammatic representation~\eqref{eq:permuations_example} by reading it from top to bottom, or more precisely by the explicit construction in App.~\ref{app:proofs} one has
$\left(\pi^{-1}\right)^\prime=\left(\pi^\prime\right)^{-1}$, i.e., the map $\pi \mapsto \pi^\prime$ commutes with taking the inverse.
In particular, we have $(\eta^{-2r})^\prime (\eta^{2r})^\prime =
\left((\eta^{2r})^\prime\right)^{-1} (\eta^{2r})^\prime = \mathrm{id} \in S_{K_{t,l}}$.
Consequently, $|(\eta^{-2r})^\prime (\eta^{2r})^\prime | = |\mathrm{id}| = |K_{t,l}| = 2(t-l)$.
Hence Eq.~\eqref{eq:matrix_elements_from_permuations} simplifies to
\begin{align}
T_{r,r} = \bra{r}\T_t^l\ket{r} = d^{2(|\eta^{2r}|_{M_{t,l}} - l)}
\end{align}
For $r=0$ one has $|\eta^{2r}|_{M_{t,l}} = |\mathrm{id}|_{M_{t,l}}=|M_{t,l}|=2l$.
Note that the matrix element $T_{0,0}$ can also be obtained from the eigenvalue equation $\T_t\ket{0}=d^2\ket{0}$.
In contrast   for $0<r<t-1$ and $t>2l$ the permutation $\eta^{2r}$ has no cycles completely contained in $M_{t,l}$, i.e., $|\eta^{2r}|_{M_{t,l}} = 0$; see App.~\ref{app:proofs}.
Minding $D_A=d^{2l}$ we thus find
\begin{align}
    T_{0,0} = D_A \quad \text{and} \quad T_{r,r} = D_A^{-1} \quad \text{for }0<r<t
\end{align}
and hence
\begin{align}
\tr(T)=D_A + \frac{t-1}{D_A} = K_A^{\mathrm{CUE}}(t),
\end{align}
where the last equality follows from Eq.~\eqref{eq:psff_CUE}.
Combining the above equation with
the Cauchy-Schwarz inequality for the Hilbert-Schmidt scalar product and the bound~\eqref{eq:bound_W}, we obtain
\begin{align}
\left| K_A(t) - K_A^{\mathrm{CUE}}(t) \right| =
\left| \tr\left(\left[W - \mathds{1}_t\right]T\right)\right|\leq \|W - \mathds{1}_t \|_{\mathrm{HS}}\|T \|_{\mathrm{HS}} <
ctd^{-t}\|T \|_{\mathrm{HS}},
\end{align}
where the right hand side can be further estimated by $\|T \|_{\mathrm{HS}}\leq t \max_{r,s}|T_{r,s}|$. The maximum is given by $D_A$ and is obtained for $r=s=0$ as can be read of from Eq.~\eqref{eq:matrix_elements_from_permuations}.
Hence eventually we obtain
\begin{align}
\left| K_A(t) - K_A^{\mathrm{CUE}}(t) \right| < cD_At^2d^{-t} \label{eq:estimate_psff}
\end{align}
for the difference between the PSFF for dual-unitary circuits and the CUE.
Alternatively, the above estimate can be written as
\begin{align}
    K_A(t)  = K_A^{\mathrm{CUE}}(t) +  \mathcal{O}\left(t^{2}d^{-t}\right),
    \label{eq:PSFF_exact_bounds_old}
\end{align}
indicating that the PSFF in dual-unitary circuits at large times is essentially identical with the random matrix result. In practice, we find this to be the case already for much shorter times $t \gtrsim 2l$, for which the right hand side of the inequliaty~\eqref{eq:estimate_psff} is not necessarily small, see Sec.~\ref{sec:numerics}  below.
The random matrix PSFF at sufficiently large times, once again, highlights the quantum chaotic nature of the dual-unitary circuits studied in our work and represents our second main result.

\begin{figure}
    \centering
    \includegraphics[width=\linewidth]{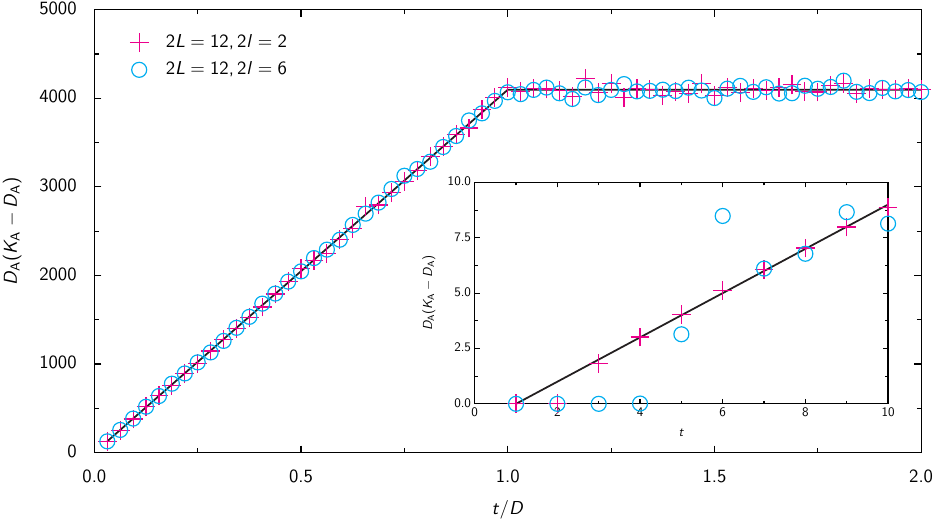}
    \caption{PSFF $K_A(t)$ for dual-unitary circuits with $d=2$, $2L=12$ and $J=0.2$ for subsystem size $2l=6$ (red crosses) and $2l=2$ (blue circles). The black line corresponds to the random matrix result $K_A^{\mathrm{CUE}}(t;D)$ with only the leading terms in $1/D$ kept. Only times which are a multiple of 64 are shown. Time is measured in units of Heisenberg time $D$ and the vertical axis is shifted and rescaled in order to compare different subsystem sizes. The inset shows the corresponding initial time regime without rescaling time by $D$ and for all times.
    Results are obtained from averaging over at least $10^4$
realizations. }
    \label{fig:PSFF_finite_size}
\end{figure}

\section{Comparison with Numerical Experiments}
\label{sec:numerics}

In this section we compare the PSFF obtained by numerical means with the random matrix result for both finite systems and directly in the thermodynamic limit.
This allows for accessing the regimes not covered by our analytical results, namely the late time plateau for finite systems and the transient time scale $l < t < 2l$ in the thermodynamic limit.

\subsection{Finite Systems}

In finite systems, numerical results are obtained by sampling from the ensemble and computing $\U_A(t) = \tr_{\Acompl}\left(\U^t\right)$ and subsequently $\tr_A\left(\U_A(t)\U_A(-t)\right)$ directly from the definition.
The results for a qubit chain of length $2L=12$ are shown in Fig.~\ref{fig:PSFF_finite_size}, where we rescale the PSFF to allow for comparing different subsystem sizes.
As predicted by Eq.~\eqref{eq:constant_psff_initial_times}, we find $K_A(t; L) = D_A$ for $t\leq l$. This is shown in the inset, which indicates that the statement even holds up to time $t=l+1$.
Between $t=l$ and  $t \sim 2l$ the PSFF quickly approaches the random matrix result. 
Keeping only the leading terms in $1/D=2^{-L}$, the latter reads $K_A^{\mathrm{CUE}}(t, D)=D_A - 1/D_A + K^{\mathrm{CUE}}(t, D)/D_A$ with $K^{\mathrm{CUE}}(t, D)$ given by Eq.~\eqref{eq:sff_CUE}.
In particular the numerically obtained PSFF exhibits the characteristic linear ramp $\sim t/D_A$ for times larger than $t=2l$ but smaller than the Heisenberg time $t=D$, inherited from the random matrix SFF.
At Heisenberg time the numerically obtained PSFF shows a sharp transition from the linear ramp into a subsequent plateau, as predicted by the random matrix result.
This agreement between numerical results and the random matrix prediction holds both for very small subsystems ($2l=2$, red crosses) and subsystems which span half the chain ($2l=6$, blue circles).
Deviations between the circuit and random matrix theory after the initial plateau
seem to originate from the finite sample size of $10^4$ realizations. Similar deviations can be observed also for the SFF, which is known to be exponentially distributed \cite{Pra1997}.
Therefore it seems reasonable that the PSFF shows similar residual fluctuations in the numerical data.
Even though these fluctuations are particularly pronounced at post Heisenberg times $t > D$, the PSFF clearly shows the expected plateau at late times. In particular upon time averaging and by Eq.~\eqref{eq:psff_averaged} this indicates that the average purity of eigenstates in dual-unitary circuits is given by the purity of random states.

\begin{figure}
    \centering
    \includegraphics[width=\linewidth]{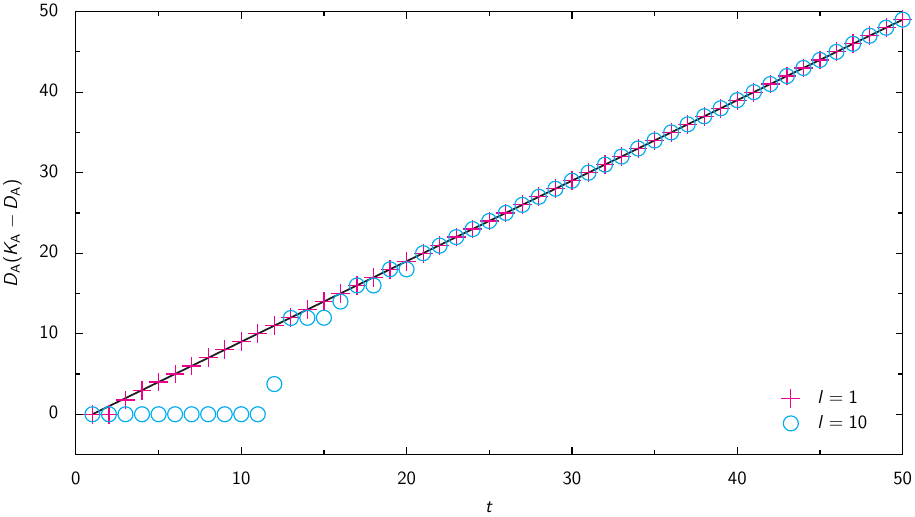}
    \caption{Rescaled PSFF for dual-unitary circuits with $d=2$ in the thermodynamic limit for $l=1$ (red crosses) and $l=10$ (blue circles). The random matrix result $K_A^{\mathrm{CUE}}(t)$ is depicted as a black line.}
    \label{fig:PSFF_td_limit}
\end{figure}

\subsection{Thermodynamic Limit}

The construction of {Secs.~\ref{sec:PSFF_td_limit}~and~\ref{sec:psff_from_permutations} allows for computing the PSFF directly in the thermodynamic limit. This is achieved by constructing the matrix $W$ by numerically inverting the Gram matrix $G$ and by computing the matrix $T$ numerically from the description via permutations.
Thus we evaluate Eq.~\eqref{eq:matrix_elements_from_permuations} by constructing the permutations $\left(\eta^{2r}\right)^\prime$ acting in the subsystem $K_{t,l}$ and by evaluating the number of cycles  $|\eta^{2r}|_{M_{t,l}}$ as well as $|\left(\eta^{-2s}\right)^\prime \left(\eta^{2r}\right)^\prime|$ numerically.
As $T$ and $W$ are only of dimension $t \times t$ this can be done for relatively large times.
We depict the resulting PSFF in Fig.~\ref{fig:PSFF_td_limit} for the minimal case $l=1$ as well as for $l=10$. To allow for comparing different subsystem sizes we again rescale the PSFF as in the finite size case in the previous section.
For initial times the exact results are provided in Sec.~\ref{sec:initial_times} and imply constant $K_A(t)=D_A$ for $t \leq l$, as it is also seen in the numerics.
Again this seems to be true even for $t \leq l+1$.

For transient times $l\leq t \leq 2l$ the PSFF quickly approaches the random matrix result as seen in the case $l=10$.
There are two sources of deviations from the random matrix result in this time regime.
The first one is is due to the periodic shifts $\eta_{2r}$ exhibiting cycles completely contained in $M_{t,l}$ yielding non-zero contributions to the exponent in Eq.~\eqref{eq:matrix_elements_from_permuations} for both diagonal and off-diagonal elements.
The second reason for deviations is rooted in the shifts $\ket{r}$ not being orthogonal and hence the Weingarten matrix not being exactly identity.
This yields contributions from the off-diagonal matrix elements of $T$ to the PSFF, which, as off-diagonal elements of the Weingarten matrix $W$ are smaller than zero, enter with a negative sign.
Ultimately, this leads to the PSFF for the circuit being smaller than for the CUE.

For larger times, $t\geq 2l$, it is only the non-orthogonality of the shifts $\ket{r}$ and the resulting deviations of the Weingarten matrix from identity, which lead to deviations. For the example shown in Fig.~\ref{fig:PSFF_td_limit} with $l=10$ the deviations  at $t=2l=20$  are remarkably small, i.e., much smaller than $10^{-5}$ and hardly visible on the shown scale.
In contrast, the right hand side of the estimate~\eqref{eq:estimate_psff} is of the order $10^{-4}D_A \sim 10^2$, highlighting that the bound~\eqref{eq:bound_W} is not tight.
For even larger times, $t\gtrsim 40$ in our example, the analytically obtained
exponential bounds in $t$ render the numerical result practically
indistinguishable from the random matrix results.

\section{Summary and Outlook}
\label{sec:outlook}

In this paper we calculate correlations between eigenstates in terms of the PSFF for dual-unitary quantum circuits exactly and compare with the corresponding random matrix result. For initial times, shorter than the considered subsystem's size, we obtain a constant PSFF as consequence of spatial locality in stark contrast to random matrix theory.
For larger times, however, the PSFF quickly approaches the random matrix result, signaling the quantum chaotic nature of the circuits.
We establish this by deriving exponentially tight bounds for the deviations
of the PSFF from the random matrix result.
While we consider systems which lack time-reversal invariance and which exhibit periodic boundary conditions, we expect our methods to be applicable also for open boundary conditions and for the time reversal invariant case.

The PSFF considered here depends, in addition to spectral correlations, also on correlations between pairs of eigenstates. As indicated in Sec.~\ref{sec:psff} higher moments of the PSFF, Eq.~\eqref{eq:psff_moments}, encode correlations between multiple eigenstates. Correspondingly, time averaged moments of the PSFF yield higher moments of the eigenstates' reduced density matrices and their respective distribution.
The general idea outlined in this work, should also allow for obtaining such moments of the PSFF by contracting more intricate tensor networks composed of multiple, possibly connected replicas.
In the same spirit, one might ask for similar dynamical quantities as the PSFF, which correspond to other measures of entanglement, e.g., mutual information.
These generalizations, however, are left for future research.

\paragraph*{Data Availability}

Figures, numerical data, and source code are available in Zenodo \cite{zenodo_data}.

\paragraph*{Acknowledgements} We thank Pavel Kos for valuable comments on the manuscript.
FF has received funding from Deutsche Forschungsgemeinschaft (DFG), Project No. 453812159 and from the European Union's Horizon Europe program under the Marie Sk{\l}odowska Curie Action GETQuantum (Grant No. 101146632). MK and AB acknowledge support from Deutsche Forschungsgemeinschaft (DFG), Project No. 497038782. 

\FloatBarrier

\begin{appendix}

\section{Partial Spectral Form Factor for Non-Chaotic Systems}
\label{app:psff_integrable}

Here we derive the PSFF for Poissonian random matrices expected to model, e.g., integrable or localized systems.
For a first model, we consider a bipartite system with Hilbert space $\Hil = \Hil_A \otimes \Hil_{\Acompl}$ of dimension $D=D_A D_{\Acompl}$ and denote the canonical product basis with $\ket{ij}$ with $i \in \{1,\ldots, D_A\}$ and $j \in \{1, \ldots, D_{\Acompl} \}$.
We define a random matrix ensemble by choosing diagonal matrices in the product basis with independent diagonal elements  uniformly distributed over the unit circle. More precisely, we choose
\begin{align}
   \U = \sum_{i=1}^{D_A}\sum_{j=1}^{D_{\Acompl}} \ue^{\ui \varphi_{ij} t}\ket{ij}\!\bra{ij}
\end{align}
with independent uniformly distributed phases $\varphi_{ij} \in \left(-\pi, \pi \right]$.
Consequently, we have
\begin{align}
\U_A(t) = \tr_{\Acompl}\left(\U^t\right) = \sum_{i=1}^{D_A}\left(\sum_{j=1}^{D_{\Acompl}} \ue^{\ui \varphi_{ij} t} \right)\ket{i}\!\bra{i}
\end{align}
and therefore
\begin{align}
\U_A(t)\U_A(-t) = \sum_{i=1}^{D_A}\left(\sum_{j,k=1}^{D_{\Acompl}} \ue^{\ui \left(\varphi_{ij} - \varphi_{ik}\right) t} \right)\ket{i}\!\bra{i} = D_{\Acompl}\mathds{1}_{D_A} +
\sum_{i=1}^{D_A}\left(\sum_{j\neq k} \ue^{\ui \left(\varphi_{ij} - \varphi_{ik}\right) t} \right)\ket{i}\!\bra{i}.
\end{align}
This yields
\begin{align}
\tr_A\left(\U_A(t)\U_A(-t)\right) = D_{\Acompl}D_A +
\sum_{i=1}^{D_A}\left(\sum_{j\neq k} \ue^{\ui \left(\varphi_{ij} - \varphi_{ik}\right) t} \right).
\end{align}
As in the second term all phases are independent from each other the average over the phases factorizes as
\begin{align}
\mathds{E}\left(\ue^{\ui \left(\varphi_{ij} - \varphi_{ik}\right) t }\right) =
    \mathds{E}\left(\ue^{\ui \varphi_{ij} t} \right)\mathds{E}\left(\ue^{-\ui \varphi_{ik} t} \right) = 0
\end{align}
with each individual average giving 0.
Ultimately, this gives
\begin{align}
    K_A^{\mathrm{Poisson}}(t; D)=D_{\Acompl}D_A = D
\end{align}
as stated in the main text with the constant given by the total Hilbert space dimension $D$. \\

As a second model we replace the fixed product eigenstates from above by random states, while keeping the same model for the eigenvalues. More precisely, we take
\begin{align}
    \U_0 = \mathrm{diagonal}\left(\ue^{\ui \varphi_1}, \ldots, \ue^{\ui \varphi_D}\right) \quad \mathrm{and} \quad
    \U = \mathcal{V}\U_0\mathcal{V}^\dagger
\end{align}
with $\mathrm{CUE}(D)$ random matrices $\mathcal{V}$ independent from the phases $\varphi_i$.
The latter are again taken to be independent and uniformly distributed in $\left(-\pi, \pi\right]$.
In this case the statistical properties of the eigenstates are the same as for the CUE and hence we might directly use Eq.~\eqref{eq:psff_CUE_finiteD} with the CUE SFF replaced by the Poissonian one with $K(t; D)=D$.
In the limit of both $D_A$ and $D_{\Acompl}$ being large, the resulting PSFF reads
\begin{align}
K_A^{\mathrm{Poisson}}(t;D)= D_A + D_{\Acompl},
\end{align}
which is again independent from time. This corresponds also to the late time plateau of the PSFF for the CUE, see Fig.~\ref{fig:intro}, but lacks the initial linear ramp.

\section{Weingarten and Gram Matrix}
\label{app:weingarten_matrix}

In this section we provide some details on the computation of the Weingarten matrix, the Gram matrix and the relevant estimates.

\subsection{Estimates of $G - \mathds{1}_t$ and $W - \mathds{1}_t$}

In the following we prove the relevant estimates for  $G - \mathds{1}_t$ and  $W - \mathds{1}_t$.
First, we compute the matrix elements of the Gram matrix by counting the numbers of cycles $|\eta^{2r}|$ of the shifts $\eta^{2r}$. The invariance of $\eta^{2r}$ under one-site shifts, i.e., $\eta^{2r} = \eta^{-1}\eta^{2r}\eta$ implies that all cycles of $\eta^{2r}$ are of the same length $a$. Consequently, $a|\eta^{2r}| = 2t$ and  $a = 2t/|\eta^{2r}|$ divides $2t$. On the other hand the length of a cycle of $\eta^{2r}$ is given by
\begin{align}
    a = \min\{a^\prime \in \mathds{N}_{>0} \,|\, 2ar = 0\, \mod 2t \}.
\end{align}
Thus there is an integer $b$ such that $2bt = 2ar = 4tr/|\eta^{2r}|$ and hence $b=2r/|\eta^{2r}|$ is an integer. Therefore,
$|\eta^{2r}|$ divides $2r$, i.e., it is a common divisor of $2t$ and $2r$. Minimality of $a$ requires $|\eta^{2r}|$ to be maximal, implying $|\eta^{2r}| = \gcd(2r, 2t) = 2 \gcd(r,t)$.
This fixes the overlapp $\braket{r|s}=\braket{0|s-r}$, defined by
Eq.~\eqref{eq:overlapp_permutations}, and leads to the matrix elements $G_{r,s}$
reported in Eq.~\eqref{eq:gram_matrix_elements}. \\

We now provide a bound on the Hilbert-Schmidt norm of $G - \mathds{1}_t$.
By normalization, the diagonal matrix elements of $G$ are $G_{r,r}=1$.
In contrast for $r \neq s$ one has $0 < |r-s| < t$ and thus $\gcd(|r-s|,t)<t$. In other words, there is an integer $a\geq 2$ such that $t=a \gcd(|r-s|,t)$.
Consequently, $\gcd(|r-s|,t)=t/a\leq t/2$ which implies
\begin{align}
   0 < G_{r,s} = d^{-2(t - \gcd(|r-s|,t))} \leq d^{-t}.
   \label{eq:bound_G_elements}
\end{align}
This implies, that $G$ is close to the identity $\mathds{1}_t$.
More precisely, the Hilbert-Schmidt norm of the real symmetric matrix $G - \mathds{1}_t$ can be obtained from
\begin{align}
   \|G - \mathds{1}_t \|_{\mathrm{HS}}^2 = \tr\left(\left[G - \mathds{1}_t\right]^2\right)
   = \tr\left(G^2\right) - 2\tr\left(G\right) + t = \tr\left(G^2\right) - t,
\end{align}
where we used $\tr(G)=t$.
For the first term we find
\begin{align}
\tr\left(G^2\right) = \sum_{r,s=0}^{t-1}G_{r,s}G_{s,r} = \sum_{r=0}^{t-1}G_{r,r}G_{r,r} + \sum_{r \neq s}G_{r,s}G_{s,r} \leq t + \sum_{r \neq s}d^{-t} < t + t^2 d ^{-2t}
\end{align}
using $G_{r,r}=1$ and the bound~\eqref{eq:bound_G_elements}.
This yields
\begin{align}
\|G - \mathds{1}_t \|_{\mathrm{HS}} < t d ^{-t}.
\end{align}
As $G$ is close to the identity, it is invertible for sufficiently large $t$, i.e. such that $td^{-t}<1$, and its inverse is given by the converging geometric series
\begin{align}
W = G^{-1} = \left(\mathds{1}_t - (\mathds{1}_t - G) \right)^{-1} = \sum_{n=0}^{\infty}(\mathds{1}_t - G)^n =  \mathds{1}_t + \sum_{n=1}^{\infty}(\mathds{1}_t - G)^n.
\end{align}
For the sum in the right hand side the triangle inequality and the 
submultiplicativity of the Hilbert-Schmidt norm imply
\begin{align}
\Big\|\sum_{n=1}^{\infty}(\mathds{1}_t - G)^n\Big\|_{\mathrm{HS}} \leq \sum_{n=1}^{\infty}\big\| \mathds{1}_t - G \|_{\mathrm{HS}}^n <
\sum_{n=1}^{\infty}\left(t d^{-t}\right)^n = \frac{td^{-t}}{1 - t d^{-t}} \leq ctd^{-t}
\end{align}
for $c \geq 1 +  \left(\ue \ln(2) - 1\right)^{-1} \approx 2.131$. The latter is just the maximum value of $(1 - td^{-t})^{-1}$.
Consequently, $W$ is close to the identity, i.e.
\begin{align}
\|W - \mathds{1}_t\|_{\mathrm{HS}} < ctd^{-t}
\end{align}
as stated in the main text.

\subsection{Weingarten Matrix for Prime $t$}

Here we provide an analytic expression of the Weingarten
matrix if $t$ is a sufficiently large prime number.
To this end, note that all the off-diagonal matrix elements of $G$ read $G_{r,s}=d^{-2(t-1)}=:\alpha $ as $\gcd(|r-s|,t)=1$ for $r\neq s$.
Therefore we may write
\begin{align}
    G = (1-\alpha)\mathds{1}_t + \alpha E,
\end{align}
where $E$ is the matrix with all entries equal to 1, i.e., $E_{ij}=1$. Defining $\beta = \alpha/(1 -  \alpha)$ we have
$(1 - \alpha)^{-1}G = \mathds{1}_t - \beta E$. As with $\alpha \ll 1$ also $\beta \ll 1$ for large $t$ the geometric series
\begin{align}
    W = G^{-1} = \frac{1}{1 - \alpha}\sum_{n=0}^{\infty}\beta^n E^n = \frac{1}{1 - \alpha}\left(\mathds{1}_t + \beta\sum_{n=1}^{\infty}\beta^{n-1} E^n\right)
\end{align}
converges and yields the Weingarten matrix $W$ as the inverse of $G$. By a straightforward inductive proof one finds $E^n=t^{n-1}E$ for $n\geq 1$.
This yields
\begin{align}
    W = \frac{1}{1 - \alpha}\left(\mathds{1}_t + \beta\sum_{n=0}^{\infty}(\beta t)^{n} E \right) = \frac{1}{1 - \alpha}\left(\mathds{1}_t + \frac{\beta}{1 - \beta t} E \right),
\end{align}
where we evaluated the geometric sum in the last equality.

\section{Details on the Computation with Permutations}
\label{app:proofs}

In this section we formalize the argument sketched in Eq.~\eqref{eq:permuations_example} by explicitly describing the map
$S_{L_t} \to S_{K_{t,l}}, \pi \mapsto \pi^\prime$ induced by taking the partial trace of permutations, i.e. $\tr_{M_{t,l}}\mathds{P}_\pi \propto \mathds{P}_{\pi^\prime}$. We denote this map by abuse of notation by $\T_t$ as well. After providing a rather general construction we apply this to the case of the PSFF and the shifts $\eta^{2r}$.

\subsection{The General Construction}

Let us describe the properties of the map $\T_t:S_{L_t} \to S_{K_{t,k}}$, which ultimately determines the matrix elements of $T$. To this end we consider a general setup:
Let $N$ a finite set, $K \subseteq N$ and denote by $S_N$ and $S_K$ the symmetric group on $N$ and $K$, respectively. We define a map $\tau:S_N \times K \to \mathds{N}_{>0}$ by
\begin{align}
    \tau(\pi, x) = \min\{\tau^\prime \in \mathds{N}_{>0} \,  |\, \pi^{\tau^\prime}(x) \in K \},
\end{align}
i.e., $\tau(\pi, x)$ is the minimal positive number of iterations of $\pi$ applied to $x$, such that the iterates return to $K$.
This is well defined, as each $x \in K$ is contained in a cycle of $\pi$ of length smaller than $|N|$. Consequently, the iterates of $x$ have to come back to $K$ after a finite number of iterations.
In the trivial case $K=N$ one simply has $\tau(\pi, x)=1$.
Replacing $\pi$ with $\pi^{-1}$ in the above definition and taking $y=\pi^{\tau(\pi, x)}(x)$ yields the property
\begin{align}
\tau(\pi, x) = \tau(\pi^{-1}, y).
\label{eq:inverse_property}
\end{align}
We then use the map $\tau$ to define
\begin{align}
    \T:S_N \to S_K: \pi \mapsto \T(\pi)
\end{align}
by setting
\begin{align}
\T(\pi):K \to K: x \mapsto \pi^{\tau(\pi, x)}(x).
\end{align}
To see, that this is well defined, i.e, that $\T(\pi)$ is indeed a permutation and in particularly bijective, note that
property~\eqref{eq:inverse_property} can be rewritten as $\tau(\pi, x) = \tau(\pi^{-1}, \T(\pi)(x))$. Consequently
\begin{align}
\T\left(\pi^{-1}\right)\circ \T\left(\pi\right)(x) & = \T\left(\pi^{-1}\right)\left(\pi^{\tau(\pi, x)}(x) \right) = \left(
\pi^{-1}\right)^{\tau\left(\pi^{-1}, \T(\pi)(x)\right)}\left(\pi^{\tau(\pi, x)}(x) \right) \\
 & = \pi^{-\tau(\pi, x)}\left(\pi^{\tau(\pi, x)}(x) \right) = \pi^{-\tau(\pi, x) + \tau(\pi, x)}(x) = x
\end{align}
and hence $\T(\pi)$ is a bijection with inverse given by $\T\left(\pi^{-1}\right)$.
In particular, we have $\T(\pi)^{-1} = \T\left(\pi^{-1}\right)$.

\subsection{Application to the Partial Spectral Form Factor}

The above construction formalizes the situation encountered in the computation of the PSFF, when identifying
$N= L_t = \{1,2,\ldots, 2t\}$ and $K = K_{t,k} = \{ k + 1, k + 2, \ldots, 2t - k - 1\}$  for $0\leq k \leq l$. and therefore $M = M_{t,k} = \{1, 2, \ldots, k\} \cup \{2t - k + 1, 2t-k+2, \ldots, 2t\} = N\setminus K$. Writing moreover $\pi^{\prime} = \T(\pi)$ one has $\left(\pi^{-1}\right)^\prime = \left(\pi^{\prime}\right)^{-1}$ as well as Eq.~\eqref{eq:partial_trace_permutation} with, as argued in the main text, $|\pi|_{M_{t,k}} = |\pi|_M$ the number of disjoint cycles of $\pi$ completely contained in $M$. \\

We first prove Eq.~\eqref{eq:iterative_cancelation}. Using Eq.~\eqref{eq:partial_trace_permutation} it suffices to show that for $\mu=\eta^{2r}$
\begin{align}
\left(\tilde{U}_0\right)^{\otimes t-k}\mathds{P}_{\mu^\prime} \left(\tilde{U}_0^\dagger\right)^{\otimes t-k} = \mathds{P}_{\mu^\prime},
\end{align}
where $\left(\tilde{U}_0\right)^{\otimes t-k}$ acts on the Hilbert space associated to the sites in $K$.
This can be achieved by pairwise canceling the gates $\U_0$ with their Hermitian conjugates similar to the example Eq.~\eqref{eq:cancelations_gates_shifts}. \\

On the level of the permutations this requires that $\mu^\prime$ maps $k+1$ and $k+2$ to $k+1 + 2s$ and $k+2 + 2s$ for some $s$, respectively, as this cancels $U_0^\dagger$ acting on the sites $k+1$ and $k+2$ with $U_0$ acting on the sites $k+1 + 2s$ and $k+2 + 2s$ due to dual unitarity.
Similar constraints need to be fulfilled for the other pairs of sites.
To formalize this, let us write $y=k + 2x - 1 \in K$ for some unique $x \in \{1, \ldots, t-k\}$. We aim for showing
\begin{align}
    \mu^\prime(y + 1) = \mu^\prime(y) + 1
    \label{eq:pairwise_mapping}
\end{align}
as this allows for the above cancellation of $\tilde{U}_0^\dagger$ acting at sites $y,y+1$ with $\tilde{U}_0$ at sites $\mu^\prime(y), \mu^\prime(y)+1$. \\

To prove the above equation we first show that $y \in \mu^{p}(M)$ if and only if $y+1 \in \mu^{p}(M)$ for all $p \in \mathds{Z}$.
Assume $y \in \mu^{p}(M)$ and hence $\mu^{-p}(y) = (y - 2pr)\mod 2t \in M$ but $y+1 \notin \mu^{p}(M)$, i.e. $\mu^{-p}(y+1) = (y+1 - 2kr)\, \mod 2t \notin M$. This can only be the case if $\mu^{-p}(y) = k$ and  $\mu^{-p}(y+1) = k + 1$ and consequently $k+1 = (k + 2x - 2pr)\, \mod 2t$. This however yields $1 = 2(x - pr) \mod 2t$, which is clearly false as the right hand side is even while the left hand side is odd. Hence $y+1 \in \mu^{p}(M)$. An analogous argument yields the other direction. \\

This property, however, implies that
\begin{align}
    \tau(\mu, y) = \tau(\mu, y+1)
\end{align}
and hence
\begin{align}
\mu^\prime(y) +1 = (y + 2\tau(\mu, y)r)\,\mod 2t + 1 = (y + 1 + 2\tau(\mu, y+1)r)\,\mod 2t = \mu^\prime(y + 1).
\end{align}
Here we also used  $(z \, \mod \, 2t) + 1 = [(z+1)\, \mod \, 2t]$ for $z \in K$.
This establishes Eq.~\eqref{eq:pairwise_mapping} and eventually proves Eq.~\eqref{eq:iterative_cancelation} in the main text. \\

In order to finalize the computation of the PSFF in the main text, it remains to show, that for $t>2l$ and $0<r<t$ the permutation $\eta^{2r}$ exhibits no cycles completely contained in $M$.
To see this, consider a cycle of length $a$, which necessarily obeys $a\geq 2$. We denote the elements in the cycle by $x_i$ for $i \in \{1,\ldots,a\}$ with $1\leq x_1 < x_2 < \ldots < x_k \leq 2t$. Note, that in general $x_{i+1} \neq \eta^{2r}(x_i)$.
By the shift invariance, $\eta^{2r}=\eta^{-1}\eta^{2r}\eta$, they are evenly spaced, i.e., $x_{i+1}=x_i + 2t/a$. As $2t/a\leq t< 2(t-l)$, there is at least one $x_j$ with $l < x_j \leq 2t-l$, i.e., $x_j \notin M$. Hence the cycle is not completely contained in $M$ and thus  $|\eta^{2r}|_M=0$. \\

\end{appendix}


\end{document}